\documentclass{article}
\usepackage[utf8]{inputenc}
\usepackage{authblk}
\usepackage{indentfirst}
\usepackage{hyperref}
\usepackage{longtable}
\usepackage{tabularray}
\usepackage{setspace}
\usepackage[margin=1in]{geometry}
\usepackage{graphicx}
\graphicspath{ {./figures/} }
\usepackage{subcaption}
\usepackage{amsmath}
\usepackage{lineno}
\usepackage{multirow}
\usepackage{color,soul}
\usepackage[version=3]{mhchem}

\usepackage[biblabel]{cite}

\title{Fluoride Frameworks as Potential Calcium Battery Cathodes}

\author[1]{Dereje Bekele Tekliye}
\author[1,*]{Gopalakrishnan Sai Gautam}

\affil[1]{Department of Materials Engineering, Indian Institute of Science, Bengaluru, 560012, India}
\affil[*]{Email: \href{mailto:saigautamg@iisc.ac.in}{saigautamg@iisc.ac.in}}

\date{}

\begin{document}

\maketitle

%%%%%% Abstract %%%%%%
\begin{abstract}
Calcium batteries (CBs) are potential next-generation energy storage devices, offering a promising alternative to lithium-ion batteries due to their theoretically high energy density, better safety, and lower costs associated with the natural abundance of calcium. However, the limited availability of positive electrode (cathode) materials has constrained the development of CBs so far. Given the similar ionic radii of Na\textsuperscript{+} and Ca\textsuperscript{2+}, structures that are effective at reversibly intercalating Na\textsuperscript{+} may be able to reversibly intercalate Ca\textsuperscript{2+} as well. In this context, transition metal fluorides (TMFs) exhibiting weberite and perovskite structures that are known for intercalating Na\textsuperscript{+} form an interesting set of possible CB cathode frameworks. Thus, we use first principles calculations to explore weberite and perovskite TMFs as CB cathodes, of compositions Ca\textsubscript{x}M\textsubscript{2}F\textsubscript{7} and Ca\textsubscript{x}MF\textsubscript{3}, respectively, where M = Ti, V, Cr, Mn, Fe, Co, or Ni. We systematically evaluate key cathode properties, including ground state structure, average Ca-intercalation voltage, thermodynamic stability (at 0 K), theoretical capacity, and Ca\textsuperscript{2+} migration barriers. Importantly, we identify Ca\textsubscript{x}Cr\textsubscript{2}F\textsubscript{7} and Ca\textsubscript{x}Mn\textsubscript{2}F\textsubscript{7} weberite frameworks as promising Ca-cathodes. Our study not only unveils potential CB cathodes but also paves the way for further advancement in TMF-based intercalation cathodes, diversifying the chemical space for next-generation energy storage systems.
\end{abstract}

%%%%%% Main Text %%%%%%

\section{Introduction}
The ubiquity of rechargeable Li-ion batteries (LIBs), with their high energy and power density, has emerged as the predominant energy source for numerous applications ranging from portable electronics, electric vehicles, and stationary energy storage.\cite{van2014rechargeable, whittingham2014ultimate, nykvist2015rapidly, tarascon2010lithium, larcher2015towards, cano2018batteries} However, the growing demand for LIBs across a wide variety of applications raises concern about Li (also Co and Ni) resource availability and associated supply-chain constraints\cite{olivetti2017lithium}, necessitating an exploration of alternate solutions to the LIB technology. Off-late, there is growing interest in Calcium batteries (CBs) as a potential next-generation battery technology due to: (i) the low standard reduction potential of Ca (-2.87~V versus standard hydrogen electrode, SHE) approaching that of Li (-3.04 V versus SHE), (ii) the multivalent nature of Ca\textsuperscript{2+}, which shuttles two valence electrons at a time, combined with the use of Ca-metal as anode, which can result in an increased volumetric energy density\cite{canepa2017odyssey, palacin2024roadmap, ponrouch2019multivalent, arroyo2019achievements, ponrouch2016towards, gummow2018calcium}, (iii) CBs utilize Ca, which is more abundant than Li (Ca: 4.2\% versus Li: 0.002\% of earth’s crust\cite{suess1956abundances}), potentially making CBs less expensive and more sustainable than LIBs. 

Despite the inherent advantages, the development of CBs is constrained by the need for suitable electrolytes and positive electrodes (cathodes). While stripping and plating Ca-metal at the negative electrode (anode) has encountered challenges due to electrolyte decomposition \cite{arroyo2019achievements, monti2019multivalent}, recent advances in the electrolyte design have enabled efficient Ca plating/stripping with the formation of a protective layer enhancing electrochemical performance.\cite{wang2018electrolyte, li2019towards, pu2020current, shyamsunder2019reversible} However, the lack of suitable cathode candidates, namely lack of thermodynamic and/or cyclic stability,\cite{gummow2018calcium} and poor Ca$^{2+}$ diffusion,\cite{rong2015materials} continues to hinder the development of practical CBs. 

Previous studies have employed both experimental and computational techniques to explore select chemistries as possible CB cathodes. The set of inorganic cathodes that have been reported with Ca so far include  CaMo\textsubscript{6}X\textsubscript{8} (X = S, Se, or Te),\cite{smeu2016theoretical} VOPO$_{4}$·2H$_{2}$O\cite{wang2020vopo}, Ca\textsubscript{x}V\textsubscript{2}O\textsubscript{5},\cite{gautam2015first, jeon2022bilayered, zhang2022towards, richard2023ultra, xu2019bilayered} CaV\textsubscript{2}O\textsubscript{4},\cite{lu2021searching, black2022elucidation} CaMn\textsubscript{2}O\textsubscript{4},\cite{chando2023exploring} MoO\textsubscript{3},\cite{cabello2018applicability, tojo2018electrochemical, chae2020calcium} NH\textsubscript{4}V\textsubscript{4}O\textsubscript{10},\cite{vo2018surfactant} CaCo\textsubscript{2}O\textsubscript{4},\cite{cabello2016advancing, park2021layered} NaFePO$_{4}$F,\cite{lipson2017calcium} TiS\textsubscript{2},\cite{tchitchekova2018} CaV\textsubscript{6}O\textsubscript{16}·2.8H\textsubscript{2}O,\cite{wang2022cav6o16} Prussian-blue analogues (PBA)\cite{kuperman2017high, shiga2015insertion, padigi2015potassium, tojo2016reversible, lipson2015rechargeable}, and other polyanionic frameworks.\cite{kim2020high, jeon2020reversible, xu2021new, tekliye2022exploration} However, only a select few of these compounds show reasonable electrochemical performance, with most suffering from inadequate cyclic stability, poor Ca-ion diffusion, and large volume changes during charge/discharge. For instance, VOPO\textsubscript{4}·2H\textsubscript{2}O demonstrates reasonable electrochemical performance against Ca-containing electrolytes,\cite{wang2020vopo} yet its susceptibility to proton intercalation and the undesirable effects of water in organic electrolytes pose challenges for practical cell applications.\cite{sai2016role}

Recently, Chando et al.\cite{chando2023exploring} explored the post-spinel phase of CaMn\textsubscript{2}O\textsubscript{4} as CB-cathode candidate using both experiments and density functional theory (DFT\cite{hohenberg1964inhomogeneous, kohn1965self}) calculations, and reported a low cycling capacity of 52 mAh/g at a rate of C/33. Prabakar et al.\cite{richard2023ultra} used a water-free $\beta$-phase Ca\textsubscript{0.14}V\textsubscript{2}O\textsubscript{5} as a Ca-cathode and reported a reversible capacity of $\sim$247 mAh/g. The authors stated that unlike conventional layered $\beta$-V\textsubscript{2}O\textsubscript{5} and $\delta$-Ca\textsubscript{x}V\textsubscript{2}O\textsubscript{5}·nH\textsubscript{2}O cathodes, the $\beta$-phase facilitates efficient insertion/extraction of Ca ions, contributing to improved cyclic stability and minimal dimensional changes during charge/discharge cycles.\cite{richard2023ultra}

In terms of polyanionic materials, Vaughey and co-workers\cite{kim2020high} reported two phosphates, namely sodium superionic conductor (NaSICON) (NaV\textsubscript{2}(PO\textsubscript{4})\textsubscript{3}) and olivine (FePO\textsubscript{4}), as high-voltage cathodes for room-temperature CBs. NaV\textsubscript{2}(PO\textsubscript{4})\textsubscript{3} demonstrates stable cycling with reversible intercalation of $\sim$0.6~mol of Ca$^{2+}$ at 3.2~V vs.~Ca, while FePO\textsubscript{4} exhibits reversible intercalation of $\sim$0.2~mol of Ca\textsuperscript{2+} at 2.9~V. Subsequently, Kang and co-workers\cite{xu2021new} reported a successful extraction/insertion of Ca from/into Na\textsubscript{0.5}VPO\textsubscript{4.8}F\textsubscript{0.7}, exhibiting exceptional cycling stability with 90\% capacity retention over 500 cycles and 87 mAh/g capacity. Additionally, Nazar and coworkers\cite{blanc2023phase} investigated the Ca--Na dual cation system, using experimental and computational methods to understand the phase evolution and kinetics of Ca\textsubscript{x}NaV\textsubscript{2}(PO\textsubscript{4})\textsubscript{3} that takes place during Ca\textsuperscript{2+} cycling.  The authors reported a reversible Ca\textsuperscript{2+} cycling capacity being limited to be x \(\approx 0.65\) in Ca\textsubscript{x}NaV\textsubscript{2}(PO\textsubscript{4})\textsubscript{3}, attributing the limit due to phase separation into Na-rich and Ca-rich phases. Also, the authors demonstrated that Na\textsuperscript{+} migration within the host framework facilitates neighboring Ca\textsuperscript{2+} migration, enabling reversible electrochemical activity.

With respect to recent computational identification of promising Ca-cathodes, Lu et al.\cite{lu2021searching} identified two promising Ca-cathode compositions through a DFT-based high-throughput screening considering average voltages, thermodynamic stability, and migration barriers, namely, post-spinel-CaV\textsubscript{2}O\textsubscript{4} and layered-CaNb\textsubscript{2}O\textsubscript{4}. While a subsequent experimental investigation by Palacín, Arroyo-de Dompablo and co-workers\cite{black2022elucidation} showed promise for reversible Ca intercalation in CaV\textsubscript{2}O\textsubscript{4}, further optimization of the electrode framework is imperative. Additionally, we performed an extensive screening study of NaSICON frameworks as potential Ca-cathodes.\cite{tekliye2022exploration} Specifically, we used  first principles calculations to determine the ground state structure, average intercalation voltage, 0~K thermodynamic stability, and migration barrier, using which we identified three promising candidate Ca-cathodes, namely, Ca\textsubscript{x}V\textsubscript{2}(PO\textsubscript{4})\textsubscript{3}, Ca\textsubscript{x}Mn\textsubscript{2}(SO\textsubscript{4})\textsubscript{3}, and Ca\textsubscript{x}Fe\textsubscript{2}(SO\textsubscript{4})\textsubscript{3}. Experimental realization of these predicted NaSICONs, except for Ca\textsubscript{x}V$_2$(PO$_4$)$_3$, is pending.

In the context of using anion chemistry to boost energy density of cathodes, Fluorine's high electronegativity triggers a strong inductive effect on transition-metal (TM) ions within the cathode.\cite{bralsford1960effect} As a result of this inductive effect, the electron density on the TM redox center reduces, resulting in a better response to addition or removal of electrons, or a higher voltage with the (de)intercalation of an electroactive ion.\cite{padhi1998tuning} In addition, the lower molar mass of F compared to polyanionic groups (e.g., phosphates) can enhance the gravimetric capacity of a given cathode. Thus, fluorine-based frameworks can exhibit higher voltages, on average, compared to oxide or other halide frameworks.  As a result, researchers typically utilize F\textsuperscript{-} to substitute anionic sites or polyanionic groups, completely/partially, yielding better cathode materials with enhanced properties.\cite{xu2021new, hua2021revisiting, fan2018high, ouyang2020effect, clement2020cation}

Fluorides as a chemical space are a promising, yet overlooked class of host materials as cathodes in intercalation-based electrochemical energy storage systems, especially in CBs. A few fluoride compounds, such as perovskite-type NaMF\textsubscript{3} (M = Fe, Mn, Ni, and Co), have been investigated as cathodes for sodium-ion batteries (NIBs).\cite{gocheva2009mechanochemical, kitajou2017cathode} NaFeF\textsubscript{3} showed an initial capacity of 130 mAhg\textsuperscript{-1}, yet suffered from high polarization and capacity fading, while the initial charge capacity of NaMnF\textsubscript{3}, NaNiF\textsubscript{3} and NaCoF\textsubscript{3} was limited to 40 mAhg\textsuperscript{-1}, largely attributable to side reaction as later demonstrated by Dimov et al.\cite{dimov2013transition} While Ca is known to occupy the perovskite framework among oxides,\cite{sai2020exploring, wexler2023multiple} fluorine-based perovskites have not been rigorously explored for CBs. 

Weberite-type fluorides (typically Fe-containing), known for their robust three-dimensional open framework interconnected by FeF\textsubscript{6} octahedra, have been explored computationally as NIB cathodes by Euchner et al.\cite{euchner2019unlocking} Moreover, weberites can be synthesized from fluoride precursors using topochemical wet chemistry routes, which employ lower temperatures compared to solid-state synthesis routes.\cite{dey2019topochemical, ghosh2024topochemical} Recently, Park et al.\cite{park2021weberite} investigated trigonal-type fluoride weberite (Na\textsubscript{2}Fe\textsubscript{2}F\textsubscript{7}), using experimental and computational methods. The authors demonstrated a high capacity of 184 mAh/g at C/20 (1C = 184 mA g\textsuperscript{-1}), which approaches the theoretical capacity of the weberite and reported a capacity retention over 88\% of the initial capacity at 2C after 1000 cycles along with an average operating voltage of $\sim$3.1 V. Park et al. attributed this exceptional electrochemical performance to preventing structural changes during Na (de)intercalation and facile Na\textsuperscript{+} diffusion. Note that the similarity in size between Na\textsuperscript{+} and Ca\textsuperscript{2+} (116 and 114 pm in an octahedral coordination environment surrounded by O\textsuperscript{2-}, respectively\cite{shannon1969effective, shannon1976revised}) indicates that cathodes with reversible Na-intercalation may hold promise as CB cathodes as well. Thus, fluoride frameworks, specifically with the weberite and perovskite structures, could be potential Ca-cathodes. 

In this work, we use DFT-based calculations to systematically explore the chemical space of weberite- and perovskite-type fluorides, with chemical formula of Ca\textsubscript{x}M\textsubscript{2}F\textsubscript{7} and Ca\textsubscript{x}MF\textsubscript{3}, respectively, as potential CB cathodes. Here, M represents a redox active 3\textit{d} transition metal, including Ti, V, Cr, Mn, Fe, Co, or Ni, while x represents the Ca composition within a range of $0 \leq$ x $\leq 1.5$ for weberites and $0 \leq$ x $\leq 0.5$ for perovskites. We calculate the ground state structure (including the stable Ca-vacancy arrangement), 0~K thermodynamic stability, average Ca intercalation voltage and Ca\textsuperscript{2+} migration barriers upon Ca (de)insertion in the weberite and perovskite fluorides.  Importantly, our systematic screening identifies Ca\textsubscript{x}Cr\textsubscript{2}F\textsubscript{7} and Ca\textsubscript{x}Mn\textsubscript{2}F\textsubscript{7} weberites as promising cathode candidates for CBs. Our study marks the first extensive exploration of fluoride frameworks as potential intercalation-type Ca-cathode materials, which can hopefully aid in the practical development of energy dense CBs. 

\section{Results}
\subsection{Structural Features}
\label{sec:structure}
\subsubsection{Weberites}

\begin{figure*}
 \centering
 \includegraphics[height=7.5cm]{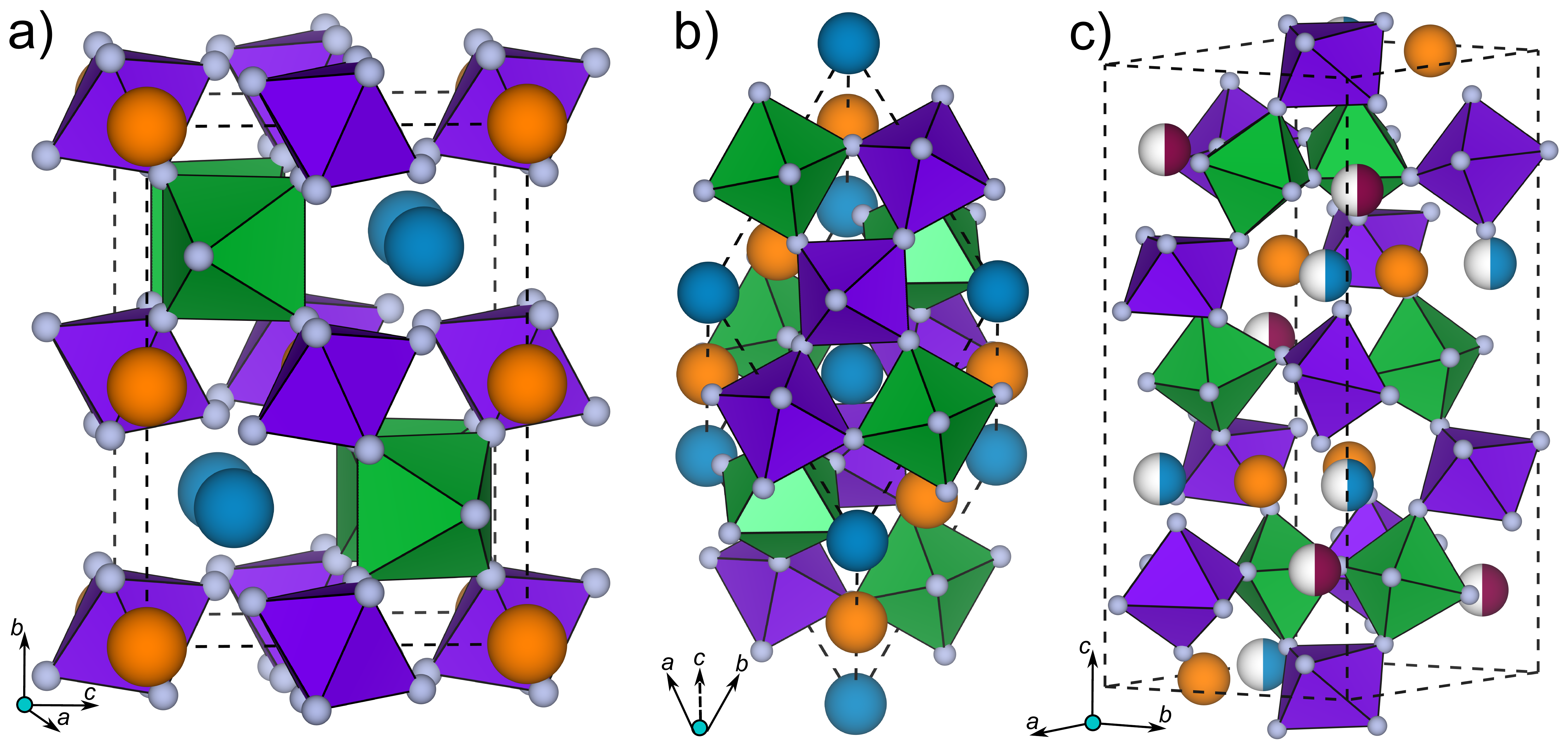}
 \caption{The crystal structure of the conventional cell (panel a) and primitive cell (panel b) of orthorhombic weberite, and conventional cell of trigonal weberite (panel c). Green and purple polyhedra indicate MF\textsubscript{6} octahedral for divalent (M\textsuperscript{2+}) and trivalent (M\textsuperscript{3+}) transition metals, respectively, where oxygen atoms are represented by grey spheres. Orange, and light blue spheres in panel a, b, c indicate Na1 and Na2 sites (equivalent to Ca1 and Ca2 in Ca-based weberites), respectively, and maroon spheres in panel c indicate Na3 sites (equivalent to Ca3). Dashed black lines indicate the extent of the unit cell.}
 \label{fig:weberites}
\end{figure*}

The weberite structure derives its name from the parent compound, namely the mineral Na\textsubscript{2}MgAlF\textsubscript{7}, and typically crystallizes in an orthorhombic structure\cite{knop1982true} with a space group of \textit{Imma} (panels a and b of Figure~{\ref{fig:weberites}}). The weberite framework is considered as a type of anion-deficient fluorite superstructure with a general stoichiometry of A\textsuperscript{(4a)}A\textquotesingle\textsuperscript{(4d)}M\textsuperscript{(4b)}M\textquotesingle\textsuperscript{(4c)}F\textsubscript{7}\textsuperscript{(4e,8h,16j)}. A and A\textquotesingle ~usually represent Na occupying two different crystallographic sites, represented by orange (Na1 or 4a sites) and blue (Na2 or 4d) spheres in Figure~{\ref{fig:weberites}}a and b. M and M\textquotesingle ~represent TM cation with a +2 and a +3 oxidation state, respectively, and are represented by green and purple octahedra in Figure~{\ref{fig:weberites}}. Note that the A and A\textquotesingle ~represent possible sites for Ca occupation in Ca-containing weberites. The conventional cell of the orthorhombic weberite (Figure~{\ref{fig:weberites}}a) consists of four A\textsubscript{2}M\textsubscript{2}F\textsubscript{7} (A = Na/Ca) formula units, with the MF\textsubscript{6} and M\textquotesingle F\textsubscript{6} octahedra sharing corners to form a three-dimensional framework. The primitive cell of the orthorhombic weberite (Figure~{\ref{fig:weberites}}b) has two A\textsubscript{2}M\textsubscript{2}F\textsubscript{7} formula units. The fluorine atoms, which are represented by grey spheres, occupy three different Wykoff positions, namely F1 (4e), F2 (8h) and F3 (16j).

Both Na1 and Na2 sites coordinate with 6+2 with neighboring fluorine atoms, with Na1 exhibiting a hexgonal bipyramidal coordination and Na2 displaying a distorted cubic (or square prism) coordination. While Na1 coordinates with all three types of F sites, Na2 coordinates with only F1 and F2. The Na2 polyhedra are interconnected via edge-sharing, resulting in a series of chains along [100], as shown in \textbf{Figure~S1a} of the supporting information (SI). On the other hand, the Na1 polyhedra share corners among each other along [010] (\textbf{Figure~S1b}), with the Na1 polyhedra also sharing some of their edges with Na2 polyhedra (\textbf{Figure~S1a}). Note that there exists another orthorhombic weberite structure, with space group \textit{Pmnb}, which is considered to be a modification of the \textit{Imma} weberite via tilting of the Na1 polyhedra resulting in a set of non-corner-sharing pentagonal bipyramidal polyhedra (\textbf{Figure~S1c}). We will not be considering the distorted \textit{Pmnb} version of the weberite structure further in our work.
    
Interestingly, the \textit{Imma} weberite framework can be considered as a repeated stacking of slabs made of (011) planes. Each slab consists of two distinct layers, namely A\textsubscript{3}M (A-rich layer) and AM\textsubscript{3} (TM-rich layer), forming an `AA' stacking sequence, as depicted in \textbf{Figure~S2a}. In turn, each layer exhibits a Kagome-like network where majority cations (A in A\textsubscript{3}M and M in AM\textsubscript{3}) form a continuous network, while minority cations occupy the Kagome ring centers as illustrated in \textbf{Figure S2b,c}. 
%Specifically, in the A\textsubscript{3}B layer, two Na1 and four Na2 sites create a hexagonal ring with trivalent cations (M) at the center, while in the AB\textsubscript{3} layer, four divalent cations (M) and two trivalent cations (M’) form a hexagonal ring with Na1 cations at the center.

In addition to the orthorhombic crystal structure, weberite can exist as a trigonal and a monoclinic polymorph, which differ in their stacking sequence of the Kagome-type layers.\cite{cai2009complex, yakubovich1993structure} \textbf{Figure~{\ref{fig:weberites}}c} depicts the conventional cell of trigonal weberite of space group \textit{P3\textsubscript{1}21} and consisting of six formula units of A\textsubscript{2}M\textsubscript{2}F\textsubscript{7}, depicted including the Wyckoff notation as   A\textsubscript{2}\textsuperscript{(6c)}M\textsuperscript{(3a)}M\textquotesingle\textsuperscript{(3b)}M\textquotesingle\textquotesingle\textsuperscript{(6c)}F\textsubscript{7}\textsuperscript{(6c)}. In trigonal weberites, there are two distinct sites for trivalent cations M/M\textquotesingle ~sitting at 3a/3b sites (purple polyhedra in Figure~{\ref{fig:weberites}}c), while the divalent cations (M\textquotesingle\textquotesingle) sit in 6c sites (green polyhedra). The TM-F polyhedra are connected to each other through corner sharing to form a three-dimensional open framework. There are 3 distinct sub-sites for Na occupation, namely Na1 (orange spheres in Figure~{\ref{fig:weberites}}c), Na2 (half-blue spheres), and Na3 (half-maroon spheres). There are 6 Na sites of each type in the conventional trigonal cell, with the Na2 and Na3 sites being half-occupied in the stoichiometric (i.e., A\textsubscript{2}M\textsubscript{2}F\textsubscript{7}) weberite. All three Na sites in the trigonal structure can be occupied by Ca, and denoted as Ca1, Ca2, and Ca3 sites in the Ca-containing structures.
%Usually, A sites are occupied by Na, and despite all Na atoms sharing the identical Wyckoff position of 6c, there are three distinct Na sub-sites, namely, Na1:6c (0.523, 0.849, 0.6653), Na2:6c (0.949, 0.186, 0.3357), and Na3:6c (0.902, 0.169, 0.8626). 

The Kagome-type stacking in trigonal weberite follows an `ABCABC' sequence, compared to the AA sequence in orthorhombic weberite, as depicted in \textbf{Figure~S3a}. The A\textsubscript{3}M layer in trigonal weberite, which is constructed with four half occupied Na2 and four fully occupied Na1 sites forming an octagonal ring with a trivalent TM at the center is shown in \textbf{Figure~S3b}. The AM\textsubscript{3} layer consists of four divalent and two trivalent TMs forming a hexagonal ring with two half-occupied Na3 sites within the ring, as displayed in \textbf{Figure~S3c}. For a point of comparison with the orthorhombic structure, the trigonal weberite with stoichiometric Na site occupancy (i.e., containing 2 Na per formula unit) is illustrated in \textbf{Figure~S4a}, and the corresponding Kagome-type stacking is depicted in \textbf{Figure~S4b-c}. The distinctions among the A-, B-, and C-type stacking layers lie in the orientation of the TMs and/or the arrangement of Na sites. For instance, Na1 sites are arranged in identical zigzag fashion along the [100], [110], and [010] directions in the A\textsubscript{3}M layer, within the A-, B-, and C-type stacking slabs, respectively (\textbf{Figure~S4b}). Note that the monoclinic weberite is reported to exist in two different Kagome-type staking sequences, namely ABAB and AABBAABB with both phases crystallizing in the \textit{C2/c} space group.\cite{grey2003structural} However, we do not include the monoclinic structure in our study since only the orthorhombic and trigonal weberites have demonstrated electrochemical activity in NIBs.{\cite{dey2019topochemical, park2021weberite}}

For Ca-containing weberites, we generated the corresponding orthorhombic (O-weberite) and trigonal (T-weberite) compositions and structures (i.e., Ca\textsubscript{x}M\textsubscript{2}F\textsubscript{7}) by substituting Ca\textsuperscript{2+} at the Na\textsuperscript{+} sites and utilising a redox-active 3$d$ TM (Ti, V, Cr, Mn, Fe, Co, or Ni) at the M sites. We considered the charged and discharged weberite compositions to be M\textsubscript{2}F\textsubscript{7} (TM oxidation state of +3.5) and Ca\textsubscript{1.5}M\textsubscript{2}F\textsubscript{7} (TM oxidation state of +2) to account for charge-neutrality of the overall composition given most TMs can reversibly access oxidation states between +2 and +4 during a redox reaction.\cite{tekliye2022exploration} We utilised the primitive cell of O-weberite and conventional cell of T-weberite to calculate the the ground state structure at both charged and discharged compositions. 
    
Based on calculated DFT total energies, we find the O-weberite to be energetically more favorable than the T-weberite at the charged compositions for V, Fe, and Co compounds, while the T-weberite is more favorable for charged Ti, Cr, Mn, and Ni compounds. In the case of discharged weberite compositions, we find the T-weberite structure to be energetically more favorable for all TMs considered. We have compiled the relative energies between O- and T-weberite structures at both the charged and discharged compositions in \textbf{Table~S2} of the SI. The occupancy of individual Ca sites in the ground state weberite configurations (i.e., T-weberite for discharged compositions), as calculated by DFT, is compiled in \textbf{Table~{\ref{tbl:Ca-va}}}, while a schematic of all ground state configurations is compiled in \textbf{Figure~S5}. Interestingly, at the discharged composition, we find Ca to partially occupy all three crystallographic sites for all TMs, indicating the lack of a strong preference for a specific site. Note that although the Ca occupancies are identical for Ti and V/Cr/Co T-weberites (\textbf{Table~{\ref{tbl:Ca-va}}}), the specific Ca-vacancy arrangements are different (\textbf{Figure~S5}).

\begin{table}
  \caption{Site occupancy of the Ca1, Ca2, and Ca3 sites in the calculated T-weberite ground states.}
  \label{tbl:Ca-va}
  \begin{tabular*}{0.48\textwidth}{@{\extracolsep{\fill}}llll}
    \hline
    \multirow{2}{*}{Compositions} & \multicolumn{3}{l}{Ca site occupancy } \\
    \cline{2-4}
    & Ca1 & Ca2 & Ca3 \\
    \hline
    Ca\textsubscript{1.5}Ti\textsubscript{2}F\textsubscript{7}  & 1/3 & 2/3 & 1/2 \\
    Ca\textsubscript{1.5}M\textsubscript{2}F\textsubscript{7}, M = V, Cr, Co  & 1/3 & 2/3 & 1/2 \\
    Ca\textsubscript{1.5}M\textsubscript{2}F\textsubscript{7}, M = Mn, Fe & 2/3 & 1/3 & 1/2 \\
    Ca\textsubscript{1.5}Ni\textsubscript{2}F\textsubscript{7} & 1/2 & 1/2 & 1/2 \\
   \hline
  \end{tabular*}
\end{table}

\subsubsection{Perovskites}
Similar to weberites, we consider perovskites of composition, Ca\textsubscript{x}MF\textsubscript{3}, obtained by inserting Ca at possible Na sites in prototypical structures. The charged and discharged compositions of Ca-containing perovskites are MF\textsubscript{3} (TM oxidation state of +3), and Ca\textsubscript{0.5}MF\textsubscript{3} (TM oxidation state of +2), respectively, which satisfies the charge neutrality constraint. Note that Na-containing perovskite-type TM fluorides exhibit different polymorphs depending on the TM. For our investigation, we considered three different perovskite polymorphs, namely cubic (\textit{Pm$\bar{3}$m} space group), orthorhombic (\textit{Pnma}), and triclinic (\textit{P$\bar{1}$}). While we used the conventional perovskite cell (containing four Ca\textsubscript{x}MF\textsubscript{3} formula units) for calculations involving orthorhombic and trigonal polymorphs, we used a $2\times2\times2$ supercell (with eight formula units) for the cubic polymorph. The set of perovskite structures we considered, along with the calculated ground state Ca-vacancy configurations are compiled in \textbf{Figure~S6}. 
    
For each TM, we considered the experimentally known ground state\cite{dimov2013transition} of the corresponding Na-containing perovskite for creating the corresponding Ca-perovskite structure. For example, we used the cubic and triclinic structures for Ca\textsubscript{x}VF\textsubscript{3} and Ca\textsubscript{x}CrF\textsubscript{3}, respectively, since the ground states of the corresponding Na-containing perovskites (i.e., NaVF\textsubscript{3} and NaCrF\textsubscript{3}) are cubic\cite{shafer1969synthesis} and triclinic\cite{bernal2020structural}. Similarly, we used the orthorhombic polymorph for Mn,\cite{ratuszna1989structure} Fe,\cite{benner1990uber} Co,\cite{yusa2012perovskite} and Ni\cite{yusa2012perovskite} containing perovskites. For the case of Ca\textsubscript{x}TiF\textsubscript{3}, we calculated the total energies of the cubic, orthorhombic, and triclinic structure with composition of Ca\textsubscript{0.5}TiF\textsubscript{3}, as the ground state of the analogous Na-containing composition is unknown. Subsequently, we found the orthorhombic structure to be energetically preferred for Ti-based perovskite (16.7 and 0.1 meV/atom lower than cubic and triclinic, respectively). For each discharged Ca\textsubscript{0.5}MF\textsubscript{3} polymorph, we constructed the corresponding charged structure by removing all Ca\textsuperscript{2+} from the discharged structure.

\subsection{Thermodynamic Stabilities}

\begin{figure*}
    \centering
    \includegraphics[height=7cm]{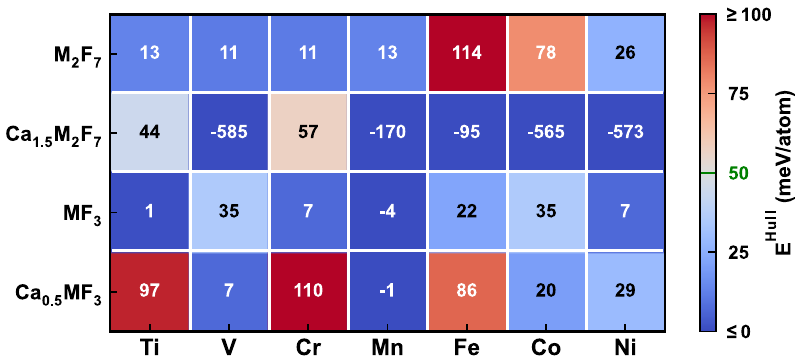}
    \caption{DFT-calculated energy above or below the ground-state convex hull (E\textsuperscript{Hull}) for charged and discharged weberites (first two rows), and perovskites (last two rows). Each column corresponds to a specific 3\textit{d} TM. Blue (red) squares indicate high degrees of stability (instability), with the specific E\textsuperscript{Hull} value of each composition listed as text annotation within each square. The green line on the legend bar indicates the rule-of-thumb E\textsuperscript{Hull}$\sim$50~meV/atom threshold for experimental synthesizability.}
    \label{fig:ther-stability}
\end{figure*}

\textbf{Figure~{\ref{fig:ther-stability}}} presents the 0~K calculated thermodynamic (in)stability, quantified using energy above or below the convex hull (E\textsuperscript{Hull}\cite{sai2020exploring}), of the charged and discharged weberite (first two rows) and perovskite (last two rows) fluorides, for all TM-containing systems considered. The E\textsuperscript{Hull} value for each composition is indicated using a text annotation in the corresponding square of \textbf{Figure~{\ref{fig:ther-stability}}}, where blue (red) squares indicate stable/metastable (unstable) compositions. Note that values of E\textsuperscript{Hull} $\geq$ 100 meV/atom (E\textsuperscript{Hull} $\leq$ 0 eV/atom) are depicted by fully red (blue) squares. Given that metastable compounds at 0~K can be synthesized under different experimental conditions, we have used a rule-of-thumb synthesizability threshold of E\textsuperscript{Hull}$\sim$50~meV/atom,\cite{sun2016thermodynamic} as indicated by the green line on the legend bar of \textbf{Figure~{\ref{fig:ther-stability}}}. Typically compositions with 0$<$E\textsuperscript{Hull}$\leq$50~meV/atom are referred to as metastable, while compositions with E\textsuperscript{Hull}$>$50~meV/atom are considered unstable. The 0~K convex hull of all ternary Ca-M-F weberites and perovskites, which comprise of possible all unary, binary, and ternary compositions that are known to exist in addition to the weberites and perovskites considered here, are compiled in \textbf{Figures S7, S8}. The compounds that are `adjacent' on the convex hull for all weberites and perovskites, i.e., compounds that thermodynamically compete with stable weberites/perovskites for existence and compounds that are the decomposition products of unstable weberites/perovskites, are compiled in \textbf{Table~S3}.
%Compounds with E\textsuperscript{Hull} $\leq$ 0 eV/atom are considered as stable phases, which lies on the convex hull with the magnitude indicating the extent of stability and the minimal energy release when the stable phase is formed from its adjacent or competing phases on the convex hull. Conversely, compounds exhibiting positive E\textsuperscript{Hull} are considered as metastable (0 $<$ E\textsuperscript{Hull} $\leq$ 50 eV/atom)/unstable (E\textsuperscript{Hull} $\geq$ 50 eV/atom) phases lying above the convex hull, where the magnitude of E\textsuperscript{Hull} quantifying the largest energy release, or the extent of metastability/instability, upon decomposition into other stable elemental, binary or ternary phases at that composition.

We observe that most of the charged compositions of weberites (first row in \textbf{Figure~{\ref{fig:ther-stability}}}) are metastable with E\textsuperscript{Hull} (in meV/atom) of 13 (Ti\textsubscript{2}F\textsubscript{7}), 11 (V\textsubscript{2}F\textsubscript{7}), 11 (Cr\textsubscript{2}F\textsubscript{7}), 13 (Mn\textsubscript{2}F\textsubscript{7}), and 26 (Ni\textsubscript{2}F\textsubscript{7}), suggesting that these compounds are potentially synthesizable. In contrast, the charged Fe- and Co-weberites are unstable, exhibiting high E\textsuperscript{Hull} of 114 and 78~meV/atom, respectively. Note that Fe\textsubscript{2}F\textsubscript{7} is thermodynamically driven to decompose into the stable adjacent compounds on the Ca-Fe-F convex hull, namely F\textsubscript{2} and FeF\textsubscript{3} (\textbf{Table S3} and \textbf{Figure~S8}). Similarly, Co\textsubscript{2}F\textsubscript{7} is expected to decompose into F\textsubscript{2} and CoF\textsubscript{3}.  

Interestingly, most of the discharged weberites (second row in \textbf{Figure~{\ref{fig:ther-stability}}}) are stable compounds with large negative E\textsuperscript{Hull} (in meV/atom) of -585 (Ca\textsubscript{1.5}V\textsubscript{2}F\textsubscript{7}), -170 (Ca\textsubscript{1.5}Mn\textsubscript{2}F\textsubscript{7}), -95 (Ca\textsubscript{1.5}Fe\textsubscript{2}F\textsubscript{7}), -565 (Ca\textsubscript{1.5}Co\textsubscript{2}F\textsubscript{7}), and -573 (Ca\textsubscript{1.5}Ni\textsubscript{2}F\textsubscript{7}), suggesting that these compositions should be experimentally synthesizable given their large thermodynamic driving force for formation. While discharged Ti-weberite is metastable (E\textsuperscript{Hull}$\sim$44~meV/atom), the discharged Cr-weberite is marginally unstable (E\textsuperscript{Hull}$\sim$57~meV/atom, marginally higher than our synthesizability threshold). Nevertheless, both Ca\textsubscript{1.5}Ti\textsubscript{2}F\textsubscript{7} and Ca\textsubscript{1.5}Cr\textsubscript{2}F\textsubscript{7} may be experimentally accessible. 

In the case of perovskites, most of the charged fluorides lie marginally above the convex hull and are likely synthesizable, including TiF\textsubscript{3}, VF\textsubscript{3}, CrF\textsubscript{3}, FeF\textsubscript{3}, CoF\textsubscript{3} and NiF\textsubscript{3}, with E\textsuperscript{Hull} (in meV/atom) of 1, 35, 7, 22, 35, and 7, respectively. Note that these charged perovskites are metastable with respect to their corresponding known trigonal trifluoride polymorphs (with space group of \textit{R$\bar{3}$cR}), as compiled in \textbf{Table~S3}.\cite{tekliye2024accuracy} The only charged perovskite that is thermodynamically stable at 0~K is MnF\textsubscript{3}, with E\textsuperscript{Hull} of -4 eV/atom. On the other hand, several discharged perovskites are unstable, with Ca\textsubscript{0.5}TiF\textsubscript{3}, Ca\textsubscript{0.5}CrF\textsubscript{3}, and Ca\textsubscript{0.5}FeF\textsubscript{3} exhibiting large E\textsuperscript{Hull} values of 97, 110, and 86~meV/atom, respectively. The metastable discharged perovskites include Ca\textsubscript{0.5}VF\textsubscript{3}, Ca\textsubscript{0.5}CoF\textsubscript{3}, and Ca\textsubscript{0.5}NiF\textsubscript{3}, with E\textsuperscript{Hull} of 7, 20, and 29~meV/atom, respectively. Similar to charged perovskites, the only discharged perovskite composition that is thermodynamically stable is the Mn-version, with E\textsuperscript{Hull} of -1~meV/atom.

Generally, it is preferable that both the charged and discharged compositions of an intercalation system are thermodynamic stable (or metastable) to prevent undesirable decomposition or conversion reactions.\cite{hannah2018balance} Thus, we identify Ca\textsubscript{x}MnF\textsubscript{3} as a potential Ca-cathode, under the constraint of both charged and discharged phases being stable. Given that there are several examples of metastable cathodes that show good electrochemical performance,\cite{sai2015intercalation, amatucci1996coo2, aurbach2000prototype, malik2013critical} we can use a constraint of both charged and discharged compositions being at most metastable for identification of candidates. In this context, we identify the following compositions to be candidate cathodes for CBs solely based on our 0~K stability calculations, namely, Ca\textsubscript{x}Ti\textsubscript{2}F\textsubscript{7}, Ca\textsubscript{x}V\textsubscript{2}F\textsubscript{7}, Ca\textsubscript{x}Cr\textsubscript{2}F\textsubscript{7}, Ca\textsubscript{x}Mn\textsubscript{2}F\textsubscript{7}, and Ca\textsubscript{x}Ni\textsubscript{2}F\textsubscript{7} within weberites and Ca\textsubscript{x}VF\textsubscript{3}, Ca\textsubscript{x}MnF\textsubscript{3}, Ca\textsubscript{x}CoF\textsubscript{3}, and Ca\textsubscript{x}NiF\textsubscript{3} among perovskites. The other compounds that we predict to be stable, such as Ca\textsubscript{1.5}Fe\textsubscript{2}F\textsubscript{7}, and Ca\textsubscript{1.5}Co\textsubscript{2}F\textsubscript{7}, may be relevant for other non-CB applications.

\subsection{Average Voltages and Theoretical Capacities}

\begin{figure}[h]
\centering
  \includegraphics[height=7cm]{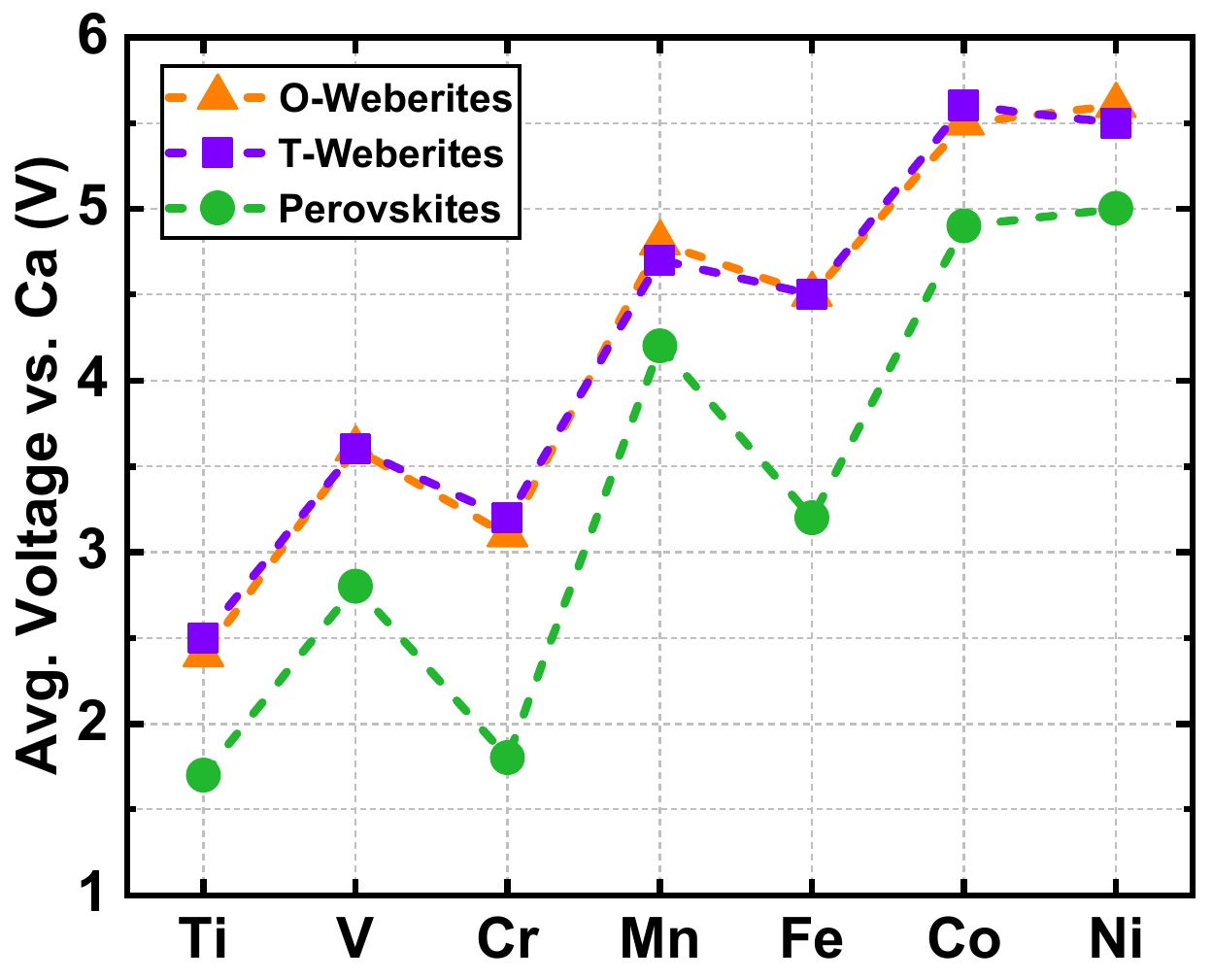}
  \caption{Computed average voltage as a function of transition metal, between the fully charged and discharged orthorhombic weberite (O-weberite, represented by orange triangles), trigonal weberite (T-weberite, purple squares) and perovskite (green circles) fluorides.}
  \label{fig:voltage}
\end{figure}

The electrochemical performance of a battery system is primarily quantified by its energy density, a metric determined by the voltage and capacity of electrodes utilised. Thus, we calculated the average Ca-intercalation voltage of the weberites and perovskites considered using DFT. Note that we computed the average voltage and theoretical capacity irrespective of thermodynamic (in)stabilities of all compositions to analyze the overall voltage-capacity trends. The calculated voltages (versus Ca) and gravimetric capacities are summarized in \textbf{Table~{\ref{tab:volt_capacity}}}. The theoretical capacities are calculated with reference to the charged compositions, i.e., MF\textsubscript{3} for perovskites and M\textsubscript{2}F\textsubscript{7} for weberites. Note that the calculated capacities decrease monotonically from Ti to Ni for both fluoride frameworks, namely from 351 to 320~mAh~g\textsuperscript{-1} in weberites, and 255 to 231~mAh~g\textsuperscript{-1} in perovskites, due to the increase in atomic mass from Ti to Ni. Also, the weberites show higher theoretical capacities compared to the perovskites, due to the higher quantity of Ca\textsuperscript{2+} (1.5 moles in weberites versus 0.5 moles in perovskites) per formula unit that can be intercalated.

\begin{table*}
     \caption{The calculated average intercalation voltage (in Volts versus Ca) and the theoretical gravimetric capacity (in mAh~g\textsuperscript{-1}) of the weberites and perovskites considered. O-Weberites and T-weberites represent voltages calculated for topotactic intercalation within the orthorhombic and trigonal weberite polymorphs, respectively. The theoretical capacities are computed with respect to the molar mass of M\textsubscript{2}F\textsubscript{7} for weberites and MF\textsubscript{3} for perovskites (M = 3$d$ TM).}
    \label{tab:volt_capacity}
    \centering
    \begin{tabular*}{\textwidth}{@{\extracolsep{\fill}}llllll} 
     \hline
     \multirow{2}{*}{M} &  &Average Voltage (V)& & \multicolumn{2}{l}{Theoretical Capacity (mAhg\textsuperscript{-1})} \\
     \cline{2-6}
     & O-Weberites & T-Weberites & Perovskites & Weberites & Perovskites \\
     \hline
     Ti & 2.4 & 2.5 & 1.7 & 351 & 255 \\
     V & 3.6 & 3.6 & 2.8 & 342 & 248 \\
     Cr & 3.1 & 3.2 & 1.8 & 339 & 246 \\
     Mn & 4.8 & 4.7 & 4.2 & 331 & 239 \\
     Fe & 4.5 & 4.5 & 3.2 & 328 & 237 \\
     Co & 5.5 & 5.6 & 4.9 & 320 & 231 \\
     Ni & 5.6 & 5.5 & 5.0 & 320 & 231 \\
     \hline    
    \end{tabular*}    
\end{table*}

\textbf{Figure~\ref{fig:voltage}} displays the calculated average intercalation voltages (in Volts versus Ca metal) for all fluoride frameworks considered here. O-weberite (orange symbols), T-weberite (purple symbols) and perovskite (green symbols) represent the calculated voltages for a topotactic intercalation within the orthorhombic weberite, trigonal weberite, and perovskite polymorphs, respectively. The average voltage is calculated across the entire Ca content range, i.e., M\textsubscript{2}F\textsubscript{7} $\leftrightarrow$ Ca\textsubscript{1.5}M\textsubscript{2}F\textsubscript{7} for weberites and MF\textsubscript{3} $\leftrightarrow$ Ca\textsubscript{0.5}MF\textsubscript{3} for perovskites. Notably, we predict the Ca intercalation voltage (in V vs. Ca) to be within the range of 2.4 (Ti) – 5.6 (Ni) for O-weberites, 2.5 V (Ti) -5.6 V (Co) for T-weberites, and 1.7 V (Ti) - 5.0 V (Ni) for perovskites. Clearly, the calculated voltages and trends for both O-weberite and T-weberite are similar across the 3\textit{d} series, suggesting that both polymorphs may exhibit similar Ca intercalation phase behavior. Additionally, both weberites display consistently higher voltages (by at least 0.5~V) than the corresponding perovskites, which predominantly originates from the charge/discharge process accessing higher TM oxidation states in weberites (+3.5) compared to perovskites (+3). 

Across the 3$d$ series, both weberites and perovskites show similar non-monotonic trends, i.e., the voltages generally increase from Ti to Ni with local minima at Cr and Fe compositions, similar to trends observed in sulfate-NaSICONs for Ca intercalation.\cite{tekliye2022exploration} In the case of T-weberites, there is a marginal dip in voltage from Co to Ni (by 0.1~V). The likely reason for the local minima in calculated voltages at Cr and Fe is the high stability of the Cr\textsuperscript{3+} and Fe\textsuperscript{3+} oxidation states, which are accessed as Ca is deintercalated. Note that Cr\textsuperscript{3+} and Fe\textsuperscript{3+} are stable due to their high-spin half-filled $t_{\textsubscript{2g}}^{\textsuperscript{3}}$ and half-filled $d$ shell ($t_{\textsubscript{2g}}^{\textsuperscript{3}} e_{\textsubscript{g}}^{\textsuperscript{2}}$) electronic configurations, respectively.\cite{vargas1996stability, shupack1991chemistry} 

In addition to the topotactic voltage, we calculated the non-topotactic average voltage for select weberites that exhibit different ground state polymorphs in their charged and discharged states. For instance, the ground state of charged weberites V\textsubscript{2}F\textsubscript{7}, Fe\textsubscript{2}F\textsubscript{7}, Co\textsubscript{2}F\textsubscript{7} are  orthorhombic, while the ground state of their corresponding discharged weberites (Ca\textsubscript{1.5}V\textsubscript{2}F\textsubscript{7}, Ca\textsubscript{1.5}Fe\textsubscript{2}F\textsubscript{7}, Ca\textsubscript{1.5}Co\textsubscript{2}F\textsubscript{7}) are the trigonal. Hence, we calculated the average voltage of V-, Fe-, and Co-based weberites between a charged orthorhombic and a discharged trigonal structure, resulting in values of 3.6, 4.5, 5.5~V vs. Ca, respectively. Importantly, the computed non-topotactic voltages are similar to the corresponding topotactic voltages ($<$0.1~V difference), suggesting insignifcant driving force for any phase transition between the trigonal and orthorhombic structures during Ca (de)intercalation. The similarity between the non-topotactic and topotactic voltages can also be rationalised via the `small' differences in the DFT-calculated total energies between the orthorhombic and trigonal structures for all TMs (see \textbf{Table~S2}). Given the marginal thermodynamic driving force for a phase transition between orthorhombic and trigonal weberites during Ca (de)intercalation, we expect the topotactic pathway to be active under electrochemical conditions. Therefore, we take into account topotactic Ca (de)intercalation in the trigonal weberite for further discussions in this manuscript, given the trigonal structure is energetically preferred at the Ca\textsubscript{1.5}M\textsubscript{2}F\textsubscript{7} composition for all TMs considered. 

Given the lower voltages with Ti-based weberites and perovskites, and the Cr-perovskite ($\leq$2.5~V), these compositions are better suited for anode applications, while the high voltages of Co- and Ni-based fluorides ($>$5~V) may be beyond the stability limits of available Ca-electrolytes. Despite this concern, it may be possible to operate the Co and Ni fluorides at a reduced voltage by extracting a lower amount of Ca from the cathode, albeit reducing the energy density. For example, the calculated voltage of 4.5~V for Fe\textsubscript{2}F\textsubscript{7} $\leftrightarrow$ Ca\textsubscript{1.5}Fe\textsubscript{2}F\textsubscript{7} decreases to 3.4~V when considering Ca\textsubscript{0.5}Fe\textsubscript{2}F\textsubscript{7} $\leftrightarrow$ Ca\textsubscript{1.5}Fe\textsubscript{2}F\textsubscript{7}, where only one mole of Ca is extracted per weberite formula unit. Thus, all weberites except Ti and all perovskites except Ti and Cr are candidate CB cathodes, based solely on our voltage calculations. 

However, including both voltage and stability data reduces the set of possible candidates. For example, the low voltages of Ti- and Cr-perovskites may appear to be suitable for anode applications, but the high thermodynamic instability of the discharged Ti- and Cr-perovskites (E\textsuperscript{Hull}$>$95~meV/atom, \textbf{Figure~{\ref{fig:ther-stability}}}) may impede any practical deployment. Similarly, Fe-based weberites and perovskites are constrained by high instabilities of the charged and discharged compositions, respectively. In contrast, Mn-weberite, Mn-perovskite, and Ni-weberite are highly promising both due to their high voltages ($>$4~V) and thermodynamic (meta)stabilities. Thus, combining our voltage and stability data, we identify weberites containing Ti, V, Cr, Mn, and Ni, and perovskites containing V, Mn, Co, and Ni to be suitable for CB electrode applications. Nevertheless, the practical feasibility of any of these identified materials to function as a CB electrode depends on their corresponding bulk Ca\textsuperscript{2+} diffusivities (\textit{vide infra}).

\subsection{Migration Barriers}
\begin{figure*}
    \centering
    \includegraphics[height=7.5cm]{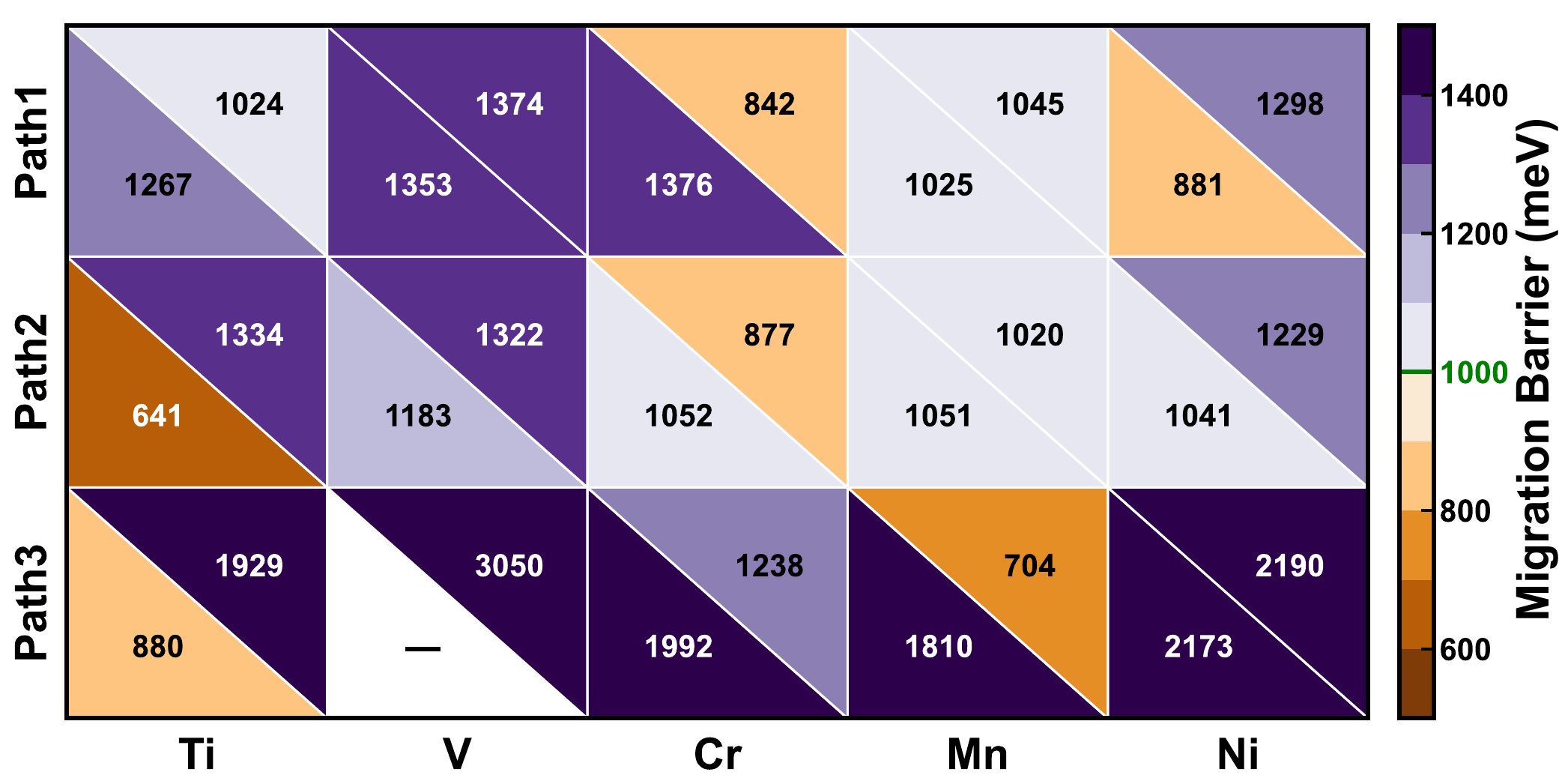}
    \caption{Calculated Ca\textsuperscript{2+} E\textsubscript{m} in the Ti-, V-, Cr-, Mn-, and Ni-weberite frameworks, indicated as text annotations. In each weberite structure, we calculated the E\textsubscript{m} across three different migration pathways, namely Path1 (top row), Path2 (middle row), and Path3 (bottom row). Bottom and top triangles within each square indicate high (charged state) and low (discharged) Ca-vacancy concentrations. Green line on the legend bar indicates the tolerance limit on E\textsubscript{m} ($\sim$1000 meV) for enabling Ca\textsubscript{2+} diffusion.  Due to convergence difficulties, we are unable to calculate a E\textsubscript{m} for V\textsubscript{2}F\textsubscript{7} along Path3 (solid black line).}
    \label{fig:barrier}
\end{figure*}

\begin{table}
    \centering
    \caption{Calculated Ca\textsuperscript{2+} E\textsubscript{m} in perovskites considered, at charged and discharged states. Solid lines represent absence of reliable E\textsubscript{m} for MnF\textsubscript{3} and Ca\textsubscript{0.5}CoF\textsubscript{3} due to convergence difficulties.}
    \label{tab:perov_barrier}
    \begin{tabular*}{0.48\textwidth}{@{\extracolsep{\fill}}lll}
    \hline
    \multirow{2}{*}{Compounds} & \multicolumn{2}{l}{Migration energy (meV)} \\
    \cline{2-3}
    & Charged & Discharged \\
    \hline
    Ca\textsubscript{x}VF\textsubscript{3} & 2875 & 1832 \\
    Ca\textsubscript{x}MnF\textsubscript{3} & \textemdash & 1980 \\
    Ca\textsubscript{x}CoF\textsubscript{3} & 1666 & \textemdash  \\ 
    Ca\textsubscript{x}NiF\textsubscript{3} & 2445 & 2120 \\
    \hline
    \end{tabular*}
\end{table}

Typically, the mobility of Ca\textsuperscript{2+} in solids is lower compared to monovalent active ions, such as Li\textsuperscript{+}, and Na\textsuperscript{+}, due to the strong electrostatic interaction of the divalent Ca\textsuperscript{2+} with the neighboring anions while migrating through a host framework.\cite{rong2015materials} Subsequently, we computed the migration barriers (E\textsubscript{m}) of Ca\textsuperscript{2+} using the nudged elastic band (NEB\cite{henkelman2000improved, sheppard2008optimization}) framework and the generalized gradient approximation (GGA\cite{perdew1996generalized}) exchange-correlation functional in the five weberite and four perovskite candidates that we had shortlisted based on our voltage and stability calculations. Note that the diffusivity of Ca\textsuperscript{2+} depends exponentially on E\textsubscript{m}, via the well-known Arrhenius relationship.\cite{gao2022design} 
    
\textbf{Figure~{\ref{fig:barrier}}} depicts the calculated E\textsubscript{m} in weberites across three different migration pathways, namely Path1 (top row), Path2 (middle row) and Path3 (bottom row) at the charged (bottom triangles) and discharged (top triangles) compositions. The individual pathways are depicted in \textbf{Figure~{\ref{fig:path}}}, wherein any of the three pathways can facilitate macroscopic Ca\textsuperscript{2+} diffusion within the three-dimensional trigonal weberite framework. The green line on the legend bar indicates the tolerance limit of $\sim$1000~meV for the E\textsubscript{m}, which corresponds to maintaining a reasonable electrochemical rate performance in a potential Ca-cathode.\cite{tekliye2022exploration} A brown (purple) triangle in \textbf{Figure {\ref{fig:barrier}}} represents a low (high) E\textsubscript{m} thereby indicating a facile (poor) Ca\textsuperscript{2+} migration. The calculated E\textsubscript{m} in perovskites are compiled in \textbf{Table~{\ref{tab:perov_barrier}}}. Due to convergence difficulties with our NEB calculations, we are unable to provide an accurate E\textsubscript{m} for the charged Ca\textsubscript{x}V\textsubscript{2}F\textsubscript{7} weberite along Path3 (solid black line). The minimum energy pathways for all weberite and perovskite structures are presented in \textbf{Figure S9} and \textbf{S10} of the SI, respectively. 

Unsurprisingly, our calculations predict fairly high Ca\textsuperscript{2+} E\textsubscript{m} for most of the weberites considered. For instance, the E\textsubscript{m} across all three paths in Ca\textsubscript{x}V\textsubscript{2}F\textsubscript{7}, at both the charged and discharged compositions are above the 1000~meV threshold, suggesting that the V-weberite is likely infeasible to be used as a CB cathode. In the case of Ti-weberite, the E\textsubscript{m} along Path1, Path2, and Path3 are within a wide range of 641 to 1929~meV. Particularly, the E\textsubscript{m} of the charged Ca\textsubscript{x}Ti\textsubscript{2}F\textsubscript{7} along Path2 (641~meV) and Path3 (880~meV), are well within the tolerance limit. However, the E\textsubscript{m} at the discharged limit are above the 1000~meV tolerance for all three paths in the discharged Ca\textsubscript{x}Ti\textsubscript{2}F\textsubscript{7} weberite, suggesting limitations with using the Ti-weberite as a CB anode. Similarly, the Ni-weberite shows reasonable E\textsubscript{m} in its charged state along Path1 (881~meV) and Path2 (1042~meV), but the E\textsubscript{m} is above the 1000~meV tolerance in the discharged state of the Ni-weberite for all pathways, making it unsuitable as a CB cathode. 

In the case of Cr- and Mn-weberites, we find at least one pathway within the corresponding structures that can facilitate Ca\textsuperscript{2+} diffusion. For instance, the E\textsubscript{m} in Ca\textsubscript{x}Cr\textsubscript{2}F\textsubscript{7} exceeds the 1000~meV threshold along both Path1 (842-1376~meV) and Path3 (1238-1992~meV), suggesting the unavailability of these pathways for Ca\textsuperscript{2+} motion. However, the E\textsubscript{m} along Path2 is in the range of 877-1052~meV, which is only marginally above the 1000~meV tolerance, suggesting that Path2 may be open for Ca\textsuperscript{2+} diffusion within the Cr-weberite. Considering the ground-state Ca-vacancy configuration of Ca\textsubscript{1.5}Cr\textsubscript{2}F\textsubscript{7} trigonal weberite (\textbf{Figure~S5d}), the fact that only Path2 remains open for Ca\textsuperscript{2+} diffusion possibly limits the amount of Ca that can be reversibly exchanged with the weberite structure to 0.83~moles per formula unit, corresponding to a capacity of $\sim$164~mAhg\textsuperscript{-1},  instead of the full 1.5~moles of Ca ($\sim$339~mAhg\textsuperscript{-1}).

In the Mn-weberite, the E\textsubscript{m} along both Path1 (1025-1045~meV) and Path2 (1020-1051~meV) are only marginally above the threshold, suggesting the feasibility of Ca\textsuperscript{2+} migration along both paths. However, Path3 is likely closed for Ca\textsuperscript{2+} motion in the Mn-weberite, since the E\textsubscript{m} in the charged state (1810~meV) is significantly above the threshold despite the low E\textsubscript{m} in the discharged state (704~meV). Given that two out of three pathways are active for Ca\textsuperscript{2+} motion in the Mn-weberite, based on the Ca-vacancy ground state configuration of Ca\textsubscript{1.5}Mn\textsubscript{2}F\textsubscript{7} (\textbf{Figure~S5g}), we conclude that the amount of Ca that can be reversibly exchanged is 1~mole per formula unit, corresponding to a capacity of $\sim$212 mAhg\textsuperscript{-1}) instead of the maximum of 1.5~moles ($\sim$331~mAhg\textsuperscript{-1}). Thus, we find both Cr-, and Mn-based weberites to be feasible to be used as CB cathodes, based on our combined set of thermodynamic stability, voltage, and Ca-mobility calculations.

With respect to perovskites, we find that the calculated E\textsubscript{m} in the V-, Mn-, Co-, and Ni-based fluorides are all significantly above the 1000~meV threshold, namely 1832-2875~meV, 1980~meV, 1666 meV, and 2120-2455~meV. Therefore, despite their promising thermodynamic stability, average voltage, and theoretical capacity, we don’t expect any of the perovskite fluorides to be CB cathode candidates due to their prohibitive Ca\textsuperscript{2+} E\textsubscript{m}. Note that we are unable to obtain reliable E\textsubscript{m} values for charged Ca\textsubscript{x}MnF\textsubscript{3} and discharged Ca\textsubscript{x}CoF\textsubscript{3} compositions, due to convergence difficulties, but we do not expect any E\textsubscript{m} that is eventually calculated for these compositions to change our conclusions. 

\section{Discussion}
\label{sec:discussion}
In this work, we used DFT-based calculations to explore the chemical space of 3$d$ TM-containing weberite and perovskite fluoride frameworks as potential intercalation cathodes for CBs. Specifically, we calculated the ground state Ca-vacancy configurations, 0~K thermodynamic stabilities, average intercalation voltages, theoretical capacities, and Ca\textsuperscript{2+} E\textsubscript{m} in several weberite and perovskite fluorides. We considered compositions of Ca\textsubscript{x}M\textsubscript{2}F\textsubscript{7} for weberites (with 0$\leq$x$\leq$1.5), and Ca\textsubscript{x}MF\textsubscript{3} for perovskites (with 0$\leq$x$\leq$0.5), where M = Ti, V, Cr, Mn, Fe, Co, or Ni. Besides obtaining several qualitative trends in stabilities, voltages, and Ca\textsuperscript{2+} mobilities, we identify two weberite frameworks, namely Cr- and Mn-based compositions, to be promising Ca-cathodes. 

While we employed the SCAN+\textit{U} functional for average voltage and thermodynamic stability calculations, we chose GGA for E\textsubscript{m} calculations, due to lower computational costs, better numerical convergence, and reliability in providing qualitative trends compared to SCAN.\cite{devi2022effect} For calculating the E\textsubscript{m} in Ca\textsubscript{1.5}V\textsubscript{2}F\textsubscript{7} along Path3 and Cr\textsubscript{2}F\textsubscript{7} along Path2, we initialized our NEB with five instead of the typical seven images since we experienced convergence difficulties with seven images. Further, in Mn\textsubscript{2}F\textsubscript{7} (along Path2), Ni\textsubscript{2}F\textsubscript{7} (all paths), Ca\textsubscript{0.5}VF\textsubscript{3} and MnF\textsubscript{3} we used a higher spring force constant of 10~eV/Å between images than the usual value of 5~eV/Å to mitigate convergence difficulties. Note that the E\textsubscript{m} usually does not change significantly with varying the spring force constant, as demonstrated by our calculations in the discharged Ca\textsubscript{x}Ni\textsubscript{2}F\textsubscript{7} system, which displays near-identical E\textsubscript{m} of 2190 and 2189~meV (along Path3) using force constants of 5 and 10~eV/Å, respectively.  
    
Recently, Long et al., demonstrated that SCAN+\textit{U} exhibits good qualitative trends in average voltages, but tends to overestimate the absolute voltage values and also the thermodynamic instabilities in Li-ion cathodes.\cite{long2021assessing} We expect similar overestimation of voltages and instabilities to extend to fluoride Ca-cathodes as well, which is why we used a slightly higher E\textsuperscript{Hull} stabilization threshold ($\sim$50 meV/atom) compared to previous studies ($\sim$25 meV/atom). Nevertheless, the majority of the fluorides considered here ($\sim$57\%) exhibit E\textsuperscript{Hull} lower than 25~meV/atom, suggesting that these fluorides should be synthesizable experimentally. 

In terms of structures considered for the fluorides, we included orthorhombic and trigonal polymorphs for weberites and cubic, orthorhombic, and triclinic polymorphs for perovskites. The choice of the polymorphs for both weberites and perovskites was largely motivated by experimental and/or electrochemical observations in the Na-systems. Another constraint influencing the choice of polymorphs is the computational cost of considering different combinations of TMs and Ca-vacancy arrangements in each structure. For instance, perovskites in general can exist in tetragonal, rhombohedral, hexagonal, and monoclinic polymorphs as well,\cite{sai2020exploring} which were not considered as possible structures here. While we hope to extend our Ca-cathode search space to other possible polymorphs in the fluoride chemical space in the future, we are optimistic that our search strategy and identification of candidates in this work are quite useful. 

Although Fe-based fluoride weberite exhibits attractive electrochemical performance as a NIB cathode,\cite{park2021weberite} we do not find the Fe-based weberite to be a feasible candidate for CB, largely due to the instability of the charged (Fe\textsubscript{2}F\textsubscript{7}) phase, which exhibits a high E\textsuperscript{Hull} of 114~meV/atom. The instability of the Fe\textsubscript{2}F\textsubscript{7} composition can be rationalised based on the instability of the 4+ oxidation state of Fe. Hence, we explored the stability of the charged Fe-weberite state with Ca already present in it such that the 4+ oxidation state is not accessed, i.e., a charged composition of Ca\textsubscript{0.5}Fe\textsubscript{2}F\textsubscript{7}. Although the E\textsuperscript{Hull} reduces to 71 meV/atom with increasing Ca content in the charged state, the Ca\textsubscript{0.5}Fe\textsubscript{2}F\textsubscript{7} phase remains above the stability threshold of $\sim$50 meV/atom, highlighting the instability of Fe-weberite even with introduction of significant residual Ca content in the structure. 
    
Given the electrochemical accessibility of the charged phase, NaFe\textsubscript{2}F\textsubscript{7} in NIBs, and the stability of the discharged phase, Ca\textsubscript{1.5}Fe\textsubscript{2}F\textsubscript{7}, in the Ca-Fe-F ternary space, insertion of Na within the Fe-weberite as a `stuffing' cation may stabilize Ca-deficient compositions and enable electrochemical exchange of Ca. This is similar to the role that Na plays in the exchange of Ca in V-based NaSICON electrodes.\cite{blanc2023phase} Also, we predict a drop in average Ca (de)intercalation voltage from 4.5~V in the Fe\textsubscript{2}F\textsubscript{7} $\leftrightarrow$ Ca\textsubscript{1.5}Fe\textsubscript{2}F\textsubscript{7} range to 3.4~V in the Ca\textsubscript{0.5}Fe\textsubscript{2}F\textsubscript{7} $\leftrightarrow$ Ca\textsubscript{1.5}Fe\textsubscript{2}F\textsubscript{7} range, which should be in line with stability windows of existing Ca electrolytes. Thus, systems containing residual Na and/or Ca in the weberite structure may be worth exploring both computationally and experimentally.  

One strategy to further explore for designing CB cathodes utilising fluoride frameworks is to explore oxyfluoride compositions, i.e., substitution of O\textsuperscript{2-} on F\textsuperscript{-} sites. Introducing a mixed anion system unlocks the potential for voltage modulation and fine-tuning other properties (such as phase behavior). Also, substituting O at F sites will likely increase the theoretical concentration of Ca\textsuperscript{2+} that can be accommodated in a weberite structure and potentially lower the E\textsubscript{m} due to slightly weaker Ca-O ionic bonds than Ca-F. Another strategy is to utilise weberite structures with multiple TMs present, similar to prior studies on Na-based weberites.\cite{euchner2019unlocking, lu2023weberite, liao2021scalable} For instance, Ni addition to the Cr-weberite, which is a candidate identified in this work, could enhance the voltage and mitigate E\textsubscript{m} of Ca\textsuperscript{2+} since the Ni-weberite exhibits high voltage and reasonable  E\textsubscript{m} along Path1 and Path2 at the charged limit.

\section{Conclusion}
As an alternative to the state-of-the-art LIBs, Ca-based electrochemical systems are a promising pathway to develop batteries with high energy density, safety, and lower cost. However, the development of practical CBs so far has been limited by the lack of suitable cathodes. Hence, we have used DFT-based calculations to systematically explore TMFs exhibiting weberite and perovskite structures, as potential CB cathodes. Specifically, we evaluated key metrics, such as thermodynamic stabilities, average Ca-intercalation voltages, theoretical capacities, and Ca\textsuperscript{2+} migration barriers, across weberites of composition Ca\textsubscript{x}M\textsubscript{2}F\textsubscript{7} and perovskites of composition Ca\textsubscript{x}MF\textsubscript{3}, where M = Ti, V, Cr, Mn, Fe, Co, or Ni. Notably, we predict a majority of the TMFs to be stable/metastable at 0~K, indicating the potential synthesizability of the TMFs considered. Our calculated average voltages exhibit higher values in weberites than perovskites, since we access higher TM oxidation states in the weberite (up to 4+) compared to perovskies (3+). Combining our stability and voltage data, we shortlisted Ti, V, Cr, Mn, and Ni weberites, and V, Mn, Co, and Ni perovskites for Ca-mobility evaluation. Subsequently, our E\textsubscript{m} calculations indicate potential for trigonal weberites, Ca\textsubscript{x}Cr\textsubscript{2}F\textsubscript{7} and Ca\textsubscript{x}Mn\textsubscript{2}F\textsubscript{7}, to be candidate CB cathodes, while none of the perovskites can facilitate Ca diffusion under reasonable electrochemical conditions.  We hope that our work will inspire further computational and experimental efforts into the usage of TMFs for CB cathodes, eventually resulting in the practical deployment of CBs.

\section{Methods}
\subsection{First Principles Calculations}
We used spin-polarized DFT\cite{hohenberg1964inhomogeneous, kohn1965self}) and projector augmented wave (PAW\cite{kresse1999ultrasoft, blochl1994improved}) potentials, as implemented in the Vienna \textit{ab initio} Simulation Package (VASP\cite{kresse1993ab, kresse1996efficient}) in all our calculations. The list of PAW potentials used in our calculations are compiled in \textbf{Table~S1} of the SI. We described the valence electrons using a plane-wave basis, expanded up to a kinetic energy cutoff of 520~eV.  We employed $\Gamma$-point-centered, Monkhorst-Pack\cite{monkhorst1976special} $k$-point meshes with a minimum of 48 subdivisions along each unit reciprocal lattice vector for geometry relaxations and a Gaussian smearing (of 0.05~eV) to integrate the Fermi surface. We used a Hubbard \textit{U} corrected strongly constrained and appropriately normed (i.e., SCAN+\textit{U}\cite{sun2015strongly, dudarev1998electron, anisimov1991band, gautam2018evaluating, long2020evaluating}) functional to describe the electronic exchange and correlation. We used \textit{U} values of (in eV) 4.5 (Ti), 4.2 (V), 1.5 (Cr), 3.8 (Mn), 5.6 (Fe), and 4.0 (Co), as derived in our previous work.\cite{tekliye2024accuracy} We utilised total energy and force convergence criteria of 10\textsuperscript{-5}~eV and \( |0.03| \) eV/Å. For all structure relaxations, we relaxed the cell shape, volume, and ionic positions, without preserving any underlying symmetry.

\subsection{Structure Generation}
We have generated crystal structures for weberites and perovskites considered in this work by including the constraint of charge neutrality. Note that stable ionic crystals are typically charge neutral. In intercalation systems that contain 3$d$ TMs, the oxidation state of the TM sets the limits on the content of the intercalant ion within the structure. Given that 3$d$ TMs can reversibly change their oxidation states from +4 to +2 during (de)intercalation, the Ca content corresponding to a +4 (+2) oxidation state represents the charged (discharged) composition of the structure. For example, the highest oxidation state that a TM can acquire in a weberite is +3.5, corresponding to a composition of Ca\textsubscript{0}M\textsubscript{2}F\textsubscript{7}, which in turn represents the charged state. Similarly, the lowest (+2) oxidation state that a TM can exhibit in a weberite is at a composition of Ca\textsubscript{1.5}M\textsubscript{2}F\textsubscript{7}, which represents the discharged state. In the case of perovskites, MF\textsubscript{3} (M\textsuperscript{3+}), and Ca\textsubscript{0.5}MF3 (M\textsuperscript{2+}) represent the charged and discharged compositions, respectively. Thus, the maximum amount of Ca that can be exchanged per formula unit of weberite and perovskite is 1.5 and 0.5~moles, which is the value we have used in our theoretical capacity calculations of \textbf{Table~{\ref{tab:volt_capacity}}}.

To generate structures for our DFT calculations, we obtained starting configurations from the international crystal structure database (ICSD\cite{hellenbrandt2004inorganic}). Specifically, we used the conventional cell of Na\textsubscript{2}Fe\textsubscript{2}F\textsubscript{7}\cite{park2021weberite} as a template for trigonal weberites, while we employed the primitive cell of Na\textsubscript{2}NiFeF\textsubscript{7}\cite{laligant1989ordered} as a template for orthorhombic weberites. We substituted Fe/Ni cites in our template structures with the relevant 3$d$ TM, while Na sites were substituted with Ca. After performing elemental substitutions, we scaled the lattice parameters of the theoretically generated structures using the ``RLSVolumePredictor"\cite{chu2018predicting} class of the pymatgen package.\cite{ong2013python} Subsequently, we used the ``enumlib" library\cite{hart2008algorithm, hart2009generating, hart2012generating, morgan2017generating} to generate symmetrically distinct Ca-vacancy arrangements within each structure.  In total, we generated 70 trigonal-Ca\textsubscript{1.5}M\textsubscript{2}F\textsubscript{7} (10 per TM), 7 trigonal-M\textsubscript{2}F\textsubscript{7} (one per TM), 14 orthorhombic-Ca\textsubscript{1.5}M\textsubscript{2}F\textsubscript{7} (2 per TM), and 7 orthorhombic-M\textsubscript{2}F\textsubscript{7} (one per TM) Ca-vacancy weberite configurations.  

Similar to weberites, we followed the elemental substitution, lattice parameter scaling, and Ca-vacancy ordering enumeration to generate Ca-containing fluoride perovskites. In total, we generated 35 configurations for Ca\textsubscript{0.5}MF\textsubscript{3} compositions, including 13 for Ti, 6 for V, 4 for Cr, and 3 each for Mn, Fe, Co, and Ni and 7 MF\textsubscript{3} (one per TM) configurations. Note that the discharged phase of perovskites exhibit different number of distinct Ca-vacancy configurations depending on the crystal structure. For instance, the orthorhombic Ca\textsubscript{0.5}MF\textsubscript{3} perovskite with (M = Ti, Mn, Fe, Co, or Ni) exhibits three possible Ca-vacancy configurations, while the cubic and triclinic versions exhibit six and four distinct configurations, respectively. We obtained a total of 13 Ca\textsubscript{0.5}TiF\textsubscript{3} configurations since we considered all three polymorphs (i.e., cubic, orthorhombic, and triclinic) for identifying the ground state configuration of the Ti-perovskite.

\subsection{Average Voltage Calculations}
In a reversible intercalation battery, where Ca\textsuperscript{2+} is inserted/extracted into/from a cathode material of composition Ca\textsubscript{y}M\textsubscript{2}F\textsubscript{7} (in a weberite structure) or Ca\textsubscript{y}MF\textsubscript{3} (in a perovskite structure), the overall redox reaction can be expressed as Eq.~\ref{eqn:v_wberite} and Eq.~\ref{eqn:v_perovskite}, respectively. 

\begin{equation}
    \ce{Ca_yM_2F_7 + (x-y) Ca^{2+} + 2(x-y) e^{-}  \rightleftharpoons Ca_xM_2F_7} \label{eqn:v_wberite}
\end{equation}

\begin{equation}
    \ce{Ca_{y}MF_{3} + (x-y) Ca^{2+} + 2(x-y) e^{-} \rightleftharpoons Ca_{x}MF_{3}} \label{eqn:v_perovskite}
\end{equation}

y and x are the concentrations of Ca in the charged and discharged weberite or perovskite fluorides, respectively. The average intercalation voltage for a given cathode material upon intercalation of (y-x) moles of Ca can be calculated using the Nernst equation (Eq.~\ref{eqn:nernst}), as below.\cite{zhou2004first}

\begin{equation}
\begin{split}
    \langle V \rangle = -\frac{\Delta G}{2(x-y)F} \\
    \approx - \frac{E(\ce{Ca_xM_{2}F_{7}/Ca_{x}MF_{3}}) - [E(\ce{Ca_{y}M_{2}F_{7}/Ca_{y}MF_{3}}) + (x-y)\mu_{\ce{Ca}}]}{2(x-y)F} 
    \label{eqn:nernst}
\end{split}
\end{equation}

$\Delta$G is the Gibbs energy change of the redox reaction, which is approximated as the total energy calculated using DFT ($\Delta G \approx \Delta E$), ignoring the $p-V$ and entropic contributions. F in Eq.~\ref{eqn:nernst} is the Faraday constant and $\mu_{Ca}$ is the Ca chemical potential in pure Ca metal (i.e., in its ground state face-centered-cubic structure).

\subsection{\textit{Ab initio} Thermodynamics}
The 0~K thermodynamic stability of the weberite and perovskite fluorides considered were evaluated with respect to the corresponding elemental, binary, and ternary compounds, as obtained from the ICSD, followed by DFT total energy calculations. For estimating stability, we exclusively considered only ordered structures that are available in the ICSD, to reduce computational complexity. In addition to the ICSD compounds, we included the theoretically obtained weberite (perovskite) compositions for estimating the stability of perovskite (weberite) compositions, since both weberites and perovskites occupy the same Ca-TM-F ternary chemical space. We used the pymatgen\cite{ong2013python} package to construct the 0~K convex hull and calculate the E\textsuperscript{Hull} for all Ca-TM-F ternary systems. Note that we used the SCAN functional without any $U$ corrections for total energy calculations of all unary systems (i.e., Ca, TMs, and F\textsubscript{2}).

\subsection{Migration Barrier Calculations}
\begin{figure*}
    \centering
    \includegraphics[height=6cm]{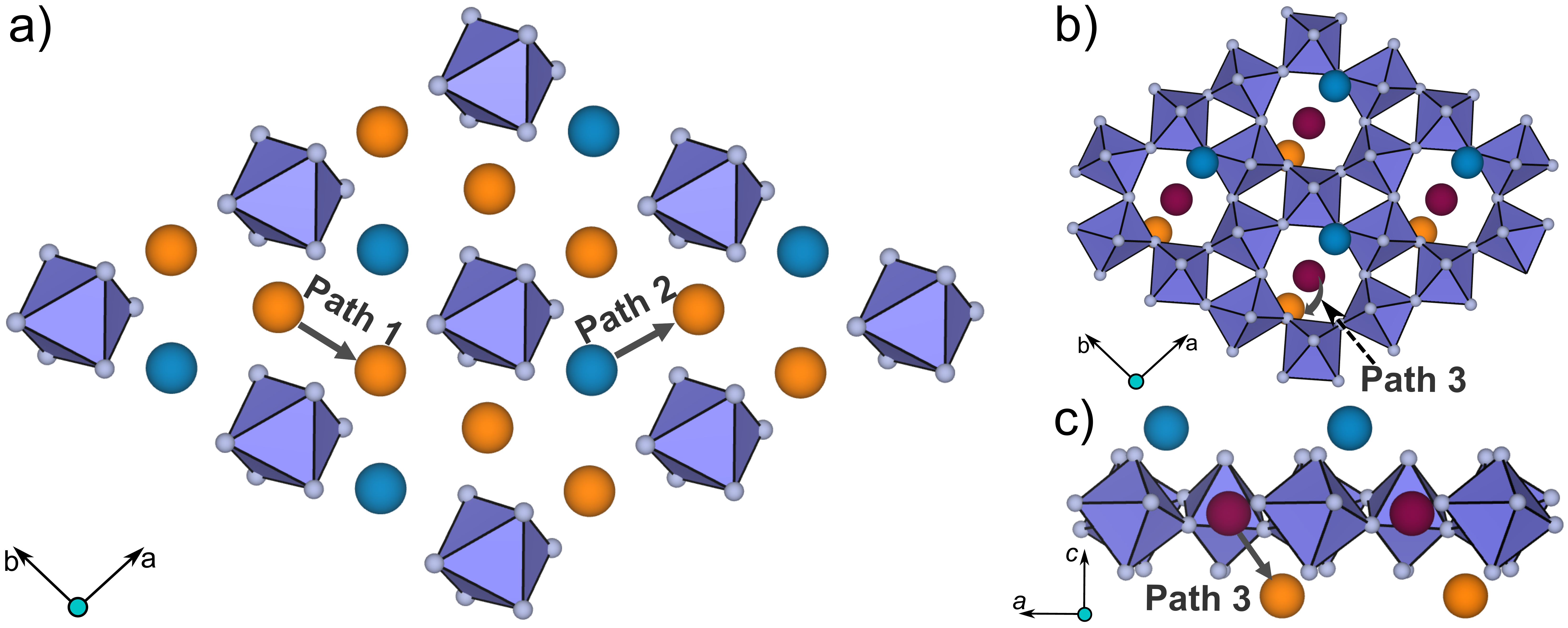}
    \caption{Ca\textsuperscript{2+} migration pathways in trigonal weberite, including Path1, Path2 (panel a), and Path3 (panels b and c). Purple polyhedra indicate MF\textsubscript{6} octahedra and F atoms are represented by grey spheres. Orange, and light blue spheres in panel a, b, c indicate Ca1 and Ca2 sites, respectively, while maroon spheres in panel b and c indicate Ca3 sites.}
    \label{fig:path}
\end{figure*}

We used DFT-based NEB\cite{henkelman2000improved, sheppard2008optimization} to evaluate the Ca\textsuperscript{2+} E\textsubscript{m} in select weberite and perovskite fluorides. We used the GGA exchange-correlation functional instead of SCAN/SCAN+\textit{U} to calculate E\textsubscript{m} to minimize computational costs and mitigate convergence difficulties.\cite{devi2022effect} For all structures, we calculated E\textsubscript{m} at both the charged (low Ca concentration) and the discharged (high Ca concentration) states. We considered a vacancy-mediated Ca\textsuperscript{2+ }migration mechanism in all structures. We converged the atomic forces to \( |0.03| \) eV/Å within the endpoint structures of each NEB. Subsequently, we initialized the minimum energy path (MEP) of the NEB calculation by linearly interpolating the lattice vectors and atomic positions to create seven intermediate images between the endpoints, with a spring constant of 5~eV/Å between images. For select structures, we used five intermediate images and/or a spring constant of 10~eV/Å (see \textbf{Section~{\ref{sec:discussion}}}). We sampled the irreducible Brillouin zone of all endpoint and NEB calculations with $\Gamma$-centered Monkhorst-Pack\cite{monkhorst1976special} $k$-point meshes with a minimum of 32 sub-divisions across each unit reciprocal lattice vector. The NEB calculations were converged until the perpendicular component of the band force between images reduced to $<$ \( |0.05| \) eV/Å. 

Within the trigonal weberite, we identified three possible Ca\textsuperscript{2+} migration pathways with distinct local environments, namely Path1 (Ca1-Ca1), Path2 (Ca2-Ca1), and Path3 (Ca3-Ca1), as depicted in \textbf{Figure~{\ref{fig:path}}}. The same pathways for Na\textsuperscript{+} migration were explored by Lu et al.\cite{lu2023weberite} in a recent study. Note that the Na arrangement along the B-type slab (within the ABCABC stacking sequence) is equivalent to the Na arrangement in the A- and C-type slabs, but along different directions. Therefore, the type of paths considered change their direction depending on the type of slab considered. For example, Path1 is along the $a$-axis or [100] direction in the A\textsubscript{3}M layer of the B-type slab, which is analogous to Path1 along the [110] direction in the A-type and [010] direction in the C-type slabs.

Notably, Path1 and Path2 migration pathways lie on the $a-b$ plane (across all A-, B-, and C-type slabs) allowing accessibility of Ca\textsuperscript{2+} at the Ca1 and Ca2 sites (within the A\textsubscript{3}M layers), as shown in \textbf{Figure~{\ref{fig:path}}a}. Specifically, all the Ca1 sites can be accessed through Path 1, while the Ca at Ca2 sites can be accessed through Path2. The Ca\textsuperscript{2+} at Ca3 sites on the AM\textsubscript{3} layer can migrate along Path3, if both E\textsubscript{m} from Ca3 to Ca1 and Ca3 to Ca2 are reasonable. Note that the Ca1-Ca3-Ca2 chains along the $c$-axis are connected via Path3, which is independent of Path1 and Path2. In case only one E\textsubscript{m}, i.e., Ca3-Ca1 or Ca3-Ca2, is reasonable, Ca\textsuperscript{2+} at Ca3 sites can migrate to available Ca1 or Ca2 sites on the A\textsubscript{3}M layer, and subsequently use Path1 or Path2 for macroscopic diffusion. However, Lu et al.{\cite{lu2023weberite}} have shown that the MEP and E\textsubscript{m} for a Na hop from Na3 to Na1 sites are similar to a Na hop from Na3 to Na2 sites.  Given the similarity between the migration paths from Na3 (Ca3) to Na1 (Ca1) and Na2 (Ca2) sites, we evaluated the E\textsubscript{m} for Ca\textsuperscript{2+} migration from Ca3 to Ca1 sites along Path3 (see panels b and c of \textbf{Figure~{\ref{fig:path}}}). Thus, depending upon the magnitude of E\textsubscript{m}, we can conclude that either Path3 will be fully open or fully closed to Ca-diffusion.

\section*{Acknowledgments}
G.S.G. acknowledges financial support from the Indian Institute of Science (IISc) and support from the Science and Engineering Research Board (SERB) of Government of India,
under Sanction Numbers SRG/2021/000201 and IPA/2021/000007. D.B.T. acknowledges financial support from Indian Institute of Science. The authors acknowledge the computational resources provided by the Supercomputer Education and Research Centre (SERC), IISc. A portion of the calculations in this work used computational resources of the supercomputer Fugaku provided by RIKEN through the HPCI System Research Project (Project ID hp220393). We acknowledge National Supercomputing Mission (NSM) for providing computing resources of `PARAM Siddhi-AI’, under National PARAM Supercomputing Facility (NPSF), C-DAC, Pune, and the resources of `Param Utkarsh' at CDAC Knowledge Park, Bengaluru. Both PARAM Siddhi-AI and PARAM Utkarsh are implemented by CDAC and supported by the Ministry of Electronics and Information Technology (MeitY) and Department of Science and Technology (DST), Government of India. The authors gratefully acknowledge the computing time provided to them on the high-performance computer noctua1 and noctua2 at the NHR Center PC2. This was funded by the Federal Ministry of Education and Research and the state governments participating on the basis of the resolutions of the GWK for the national high-performance computing at universities (\hyperlink{www.nhr-verein.de/unsere-partner}{www.nhr-verein.de/unsere-partner}). The computations for this research were performed using computing resources under project hpc-prf-emdft.

\subsection*{Author Contributions} 
GSG: Conceptualization, Supervision, Methodology, Writing, and Editing. DBT: Data generation and curation, Visualization, Writing, and Editing. 

%\subsection*{Funding}

\subsection*{Conflicts of Interest}
There are no conflicts to declare.

\section*{Supplementary Materials}
Electronic Supporting Information is available online at , Projector augmented wave potentials used, structural features of orthorhombic and trigonal weberites, ground state Ca-vacancy configurations in both weberites and perovskites, 0~K phase diagrams, and minimum energy pathways of Ca\textsuperscript{2+} migration in select structures.

%\section*{Guidelines for References}

\bibliographystyle{unsrt}
\bibliography{jmcabibfile}

\begin{thebibliography}{100}

\bibitem{van2014rechargeable}
Noorden~R Van.
\newblock The rechargeable revolution: A better battery.
\newblock {\em Nature}, 507(7490):26--28, 2014.

\bibitem{whittingham2014ultimate}
M~Stanley Whittingham.
\newblock Ultimate limits to intercalation reactions for lithium batteries.
\newblock {\em Chemical Reviews}, 114(23):11414--11443, 2014.

\bibitem{nykvist2015rapidly}
Bj{\"o}rn Nykvist and M{\aa}ns Nilsson.
\newblock Rapidly falling costs of battery packs for electric vehicles.
\newblock {\em Nature Climate Change}, 5(4):329--332, 2015.

\bibitem{tarascon2010lithium}
Jean-Marie Tarascon.
\newblock Is lithium the new gold?
\newblock {\em Nature Chemistry}, 2(6):510--510, 2010.

\bibitem{larcher2015towards}
Dominique Larcher and Jean-Marie Tarascon.
\newblock Towards greener and more sustainable batteries for electrical energy
  storage.
\newblock {\em Nature Chemistry}, 7(1):19--29, 2015.

\bibitem{cano2018batteries}
Zachary~P Cano, Dustin Banham, Siyu Ye, Andreas Hintennach, Jun Lu, Michael
  Fowler, and Zhongwei Chen.
\newblock Batteries and fuel cells for emerging electric vehicle markets.
\newblock {\em Nature Energy}, 3(4):279--289, 2018.

\bibitem{olivetti2017lithium}
Elsa~A Olivetti, Gerbrand Ceder, Gabrielle~G Gaustad, and Xinkai Fu.
\newblock Lithium-ion battery supply chain considerations: analysis of
  potential bottlenecks in critical metals.
\newblock {\em Joule}, 1(2):229--243, 2017.

\bibitem{canepa2017odyssey}
Pieremanuele Canepa, Gopalakrishnan Sai~Gautam, Daniel~C Hannah, Rahul Malik,
  Miao Liu, Kevin~G Gallagher, Kristin~A Persson, and Gerbrand Ceder.
\newblock Odyssey of multivalent cathode materials: open questions and future
  challenges.
\newblock {\em Chemical Reviews}, 117(5):4287--4341, 2017.

\bibitem{palacin2024roadmap}
M~Rosa Palacin, Patrik Johansson, Robert Dominko, Ben Dlugatch, Doron Aurbach,
  Zhenyou Li, Maximilian Fichtner, Olivera Lu{\v{z}}anin, Jan Bitenc, Zhixuan
  Wei, et~al.
\newblock Roadmap on multivalent batteries.
\newblock {\em Journal of Physics: Energy}, 2024.

\bibitem{ponrouch2019multivalent}
Alexandre Ponrouch, Jan Bitenc, Robert Dominko, Niklas Lindahl, Patrik
  Johansson, and M~Rosa Palac{\'\i}n.
\newblock Multivalent rechargeable batteries.
\newblock {\em Energy Storage Materials}, 20:253--262, 2019.

\bibitem{arroyo2019achievements}
M~Elena Arroyo-de Dompablo, Alexandre Ponrouch, Patrik Johansson, and M~Rosa
  Palac{\'\i}n.
\newblock Achievements, challenges, and prospects of calcium batteries.
\newblock {\em Chemical Reviews}, 120(14):6331--6357, 2019.

\bibitem{ponrouch2016towards}
Alexandre Ponrouch, Carlos Frontera, Fanny Bard{\'e}, and M~Rosa Palac{\'\i}n.
\newblock Towards a calcium-based rechargeable battery.
\newblock {\em Nature Materials}, 15(2):169--172, 2016.

\bibitem{gummow2018calcium}
Rosalind~J Gummow, George Vamvounis, M~Bobby Kannan, and Yinghe He.
\newblock Calcium-ion batteries: current state-of-the-art and future
  perspectives.
\newblock {\em Advanced Materials}, 30(39):1801702, 2018.

\bibitem{suess1956abundances}
Hans~E Suess and Harold~C Urey.
\newblock Abundances of the elements.
\newblock {\em Reviews of Modern Physics}, 28(1):53, 1956.

\bibitem{monti2019multivalent}
Damien Monti, Alexandre Ponrouch, Rafael~B Araujo, Fanny Barde, Patrik
  Johansson, and M~Rosa Palac{\'\i}n.
\newblock Multivalent batteries—prospects for high energy density: Ca
  batteries.
\newblock {\em Frontiers in Chemistry}, 7:79, 2019.

\bibitem{wang2018electrolyte}
Da~Wang, Xiangwen Gao, Yuhui Chen, Liyu Jin, Christian Kuss, and Peter~G.
  Bruce.
\newblock Plating and stripping calcium in an organic electrolyte.
\newblock {\em Nature Materials}, 17(1):16--20, 2018.

\bibitem{li2019towards}
Zhenyou Li, Olaf Fuhr, Maximilian Fichtner, and Zhirong Zhao-Karger.
\newblock Towards stable and efficient electrolytes for room-temperature
  rechargeable calcium batteries.
\newblock {\em Energy \& Environmental Science}, 12(12):3496--3501, 2019.

\bibitem{pu2020current}
Shengda~D Pu, Chen Gong, Xiangwen Gao, Ziyang Ning, Sixie Yang, John-Joseph
  Marie, Boyang Liu, Robert~A House, Gareth~O Hartley, Jun Luo, et~al.
\newblock Current-density-dependent electroplating in ca electrolytes: From
  globules to dendrites.
\newblock {\em ACS Energy Letters}, 5(7):2283--2290, 2020.

\bibitem{shyamsunder2019reversible}
Abhinandan Shyamsunder, Lauren~E Blanc, Abdeljalil Assoud, and Linda~F Nazar.
\newblock Reversible calcium plating and stripping at room temperature using a
  borate salt.
\newblock {\em ACS Energy Letters}, 4(9):2271--2276, 2019.

\bibitem{rong2015materials}
Ziqin Rong, Rahul Malik, Pieremanuele Canepa, Gopalakrishnan Sai~Gautam, Miao
  Liu, Anubhav Jain, Kristin Persson, and Gerbrand Ceder.
\newblock Materials design rules for multivalent ion mobility in intercalation
  structures.
\newblock {\em Chemistry of Materials}, 27(17):6016--6021, 2015.

\bibitem{smeu2016theoretical}
Manuel Smeu, Md~Sazzad Hossain, Zi~Wang, Vladimir Timoshevskii, Kirk~H Bevan,
  and Karim Zaghib.
\newblock Theoretical investigation of chevrel phase materials for cathodes
  accommodating ca2+ ions.
\newblock {\em Journal of Power Sources}, 306:431--436, 2016.

\bibitem{wang2020vopo}
Junjun Wang, Shuangshuang Tan, Fangyu Xiong, Ruohan Yu, Peijie Wu, Lianmeng
  Cui, and Qinyou An.
\newblock Vopo 4{\textperiodcentered} 2h 2 o as a new cathode material for
  rechargeable ca-ion batteries.
\newblock {\em Chemical communications}, 56(26):3805--3808, 2020.

\bibitem{gautam2015first}
Gopalakrishnan~Sai Gautam, Pieremanuele Canepa, Rahul Malik, Miao Liu, Kristin
  Persson, and Gerbrand Ceder.
\newblock First-principles evaluation of multi-valent cation insertion into
  orthorhombic v 2 o 5.
\newblock {\em Chemical communications}, 51(71):13619--13622, 2015.

\bibitem{jeon2022bilayered}
Boosik Jeon, Hunho~H Kwak, and Seung-Tae Hong.
\newblock Bilayered ca0. 28v2o5{\textperiodcentered} h2o: High-capacity cathode
  material for rechargeable ca-ion batteries and its charge storage mechanism.
\newblock {\em Chemistry of Materials}, 34(4):1491--1498, 2022.

\bibitem{zhang2022towards}
Xiao Zhang, Xiaoming Xu, Bo~Song, Manyi Duan, Jiashen Meng, Xuanpeng Wang,
  Zhitong Xiao, Lin Xu, and Liqiang Mai.
\newblock Towards a stable layered vanadium oxide cathode for high-capacity
  calcium batteries.
\newblock {\em Small}, 18(43):2107174, 2022.

\bibitem{richard2023ultra}
Samuel~Jayaraj Richard~Prabakar, Amol~Bhairuba Ikhe, Woon-Bae Park, Docheon
  Ahn, Kee-Sun Sohn, and Myoungho Pyo.
\newblock Ultra-high capacity and cyclability of $\beta$-phase ca0. 14v2o5 as a
  promising cathode in calcium-ion batteries.
\newblock {\em Advanced Functional Materials}, 33(29):2301399, 2023.

\bibitem{xu2019bilayered}
Xiaoming Xu, Manyi Duan, Yunfan Yue, Qi~Li, Xiao Zhang, Lu~Wu, Peijie Wu,
  Bo~Song, and Liqiang Mai.
\newblock Bilayered mg0. 25v2o5{\textperiodcentered} h2o as a stable cathode
  for rechargeable ca-ion batteries.
\newblock {\em ACS Energy Letters}, 4(6):1328--1335, 2019.

\bibitem{lu2021searching}
Wang Lu, Juefan Wang, Gopalakrishnan Sai~Gautam, and Pieremanuele Canepa.
\newblock Searching ternary oxides and chalcogenides as positive electrodes for
  calcium batteries.
\newblock {\em Chemistry of Materials}, 33(14):5809--5821, 2021.

\bibitem{black2022elucidation}
Ashley~P Black, Carlos Frontera, Arturo Torres, Miguel Recio-Poo, Patrick
  Rozier, Juan~D Forero-Saboya, Fran{\c{c}}ois Fauth, Esteban Urones-Garrote,
  M~Elena Arroyo-de Dompablo, and M~Rosa Palac{\'\i}n.
\newblock Elucidation of the redox activity of ca2mno3. 5 and cav2o4 in calcium
  batteries using operando xrd: charge compensation mechanism and
  reversibility.
\newblock {\em Energy Storage Materials}, 47:354--364, 2022.

\bibitem{chando2023exploring}
Paul~Alexis Chando, Sihe Chen, Jacob~Matthew Shellhamer, Elizabeth Wall, Xinlu
  Wang, Robson Schuarca, Manuel Smeu, and Ian~Dean Hosein.
\newblock Exploring calcium manganese oxide as a promising cathode material for
  calcium-ion batteries.
\newblock {\em Chemistry of Materials}, 35(20):8371--8381, 2023.

\bibitem{cabello2018applicability}
Marta Cabello, Francisco Nacimiento, Ricardo Alc{\'a}ntara, Pedro Lavela,
  Carlos Perez~Vicente, and Jos{\'e}~L Tirado.
\newblock Applicability of molybdite as an electrode material in calcium
  batteries: a structural study of layer-type ca x moo3.
\newblock {\em Chemistry of Materials}, 30(17):5853--5861, 2018.

\bibitem{tojo2018electrochemical}
Tomohiro Tojo, Hayato Tawa, Noriyuki Oshida, Ryoji Inada, and Yoji Sakurai.
\newblock Electrochemical characterization of a layered $\alpha$-moo3 as a new
  cathode material for calcium ion batteries.
\newblock {\em Journal of Electroanalytical Chemistry}, 825:51--56, 2018.

\bibitem{chae2020calcium}
Munseok~S Chae, Hunho~H Kwak, and Seung-Tae Hong.
\newblock Calcium molybdenum bronze as a stable high-capacity cathode material
  for calcium-ion batteries.
\newblock {\em ACS Applied Energy Materials}, 3(6):5107--5112, 2020.

\bibitem{vo2018surfactant}
Thuan~Ngoc Vo, Hyeongwoo Kim, Jaehyun Hur, Wonchang Choi, and Il~Tae Kim.
\newblock Surfactant-assisted ammonium vanadium oxide as a superior cathode for
  calcium-ion batteries.
\newblock {\em Journal of materials chemistry A}, 6(45):22645--22654, 2018.

\bibitem{cabello2016advancing}
Marta Cabello, Francisco Nacimiento, Jos{\'e}~R Gonz{\'a}lez, Gregorio Ortiz,
  Ricardo Alc{\'a}ntara, Pedro Lavela, Carlos P{\'e}rez-Vicente, and Jos{\'e}~L
  Tirado.
\newblock Advancing towards a veritable calcium-ion battery: Caco2o4 positive
  electrode material.
\newblock {\em Electrochemistry Communications}, 67:59--64, 2016.

\bibitem{park2021layered}
Haesun Park, Christopher~J Bartel, Gerbrand Ceder, and Peter Zapol.
\newblock Layered transition metal oxides as ca intercalation cathodes: A
  systematic first-principles evaluation.
\newblock {\em Advanced Energy Materials}, 11(48):2101698, 2021.

\bibitem{lipson2017calcium}
Albert~L Lipson, Soojeong Kim, Baofei Pan, Chen Liao, Timothy~T Fister, and
  Brian~J Ingram.
\newblock Calcium intercalation into layered fluorinated sodium iron phosphate.
\newblock {\em Journal of Power Sources}, 369:133--137, 2017.

\bibitem{tchitchekova2018}
Deyana~S Tchitchekova, Alexandre Ponrouch, Roberta Verrelli, Thibault Broux,
  Carlos Frontera, Andrea Sorrentino, Fanny Bard{\'e}, Neven Biskup, M~Elena
  Arroyo-de Dompablo, and M~Rosa Palacin.
\newblock Electrochemical intercalation of calcium and magnesium in tis2:
  fundamental studies related to multivalent battery applications.
\newblock {\em Chemistry of Materials}, 30(3):847--856, 2018.

\bibitem{wang2022cav6o16}
Junjun Wang, Jianxiang Wang, Yalong Jiang, Fangyu Xiong, Shuangshuang Tan, Fan
  Qiao, Jinghui Chen, Qinyou An, and Liqiang Mai.
\newblock Cav6o16{\textperiodcentered} 2.8 h2o with ca2+ pillar and water
  lubrication as a high-rate and long-life cathode material for ca-ion
  batteries.
\newblock {\em Advanced Functional Materials}, 32(25):2113030, 2022.

\bibitem{kuperman2017high}
Neal Kuperman, Prasanna Padigi, Gary Goncher, David Evans, Joseph Thiebes, and
  Raj Solanki.
\newblock High performance prussian blue cathode for nonaqueous ca-ion
  intercalation battery.
\newblock {\em Journal of Power Sources}, 342:414--418, 2017.

\bibitem{shiga2015insertion}
Tohru Shiga, Hiroki Kondo, Yuichi Kato, and Masae Inoue.
\newblock Insertion of calcium ion into prussian blue analogue in nonaqueous
  solutions and its application to a rechargeable battery with dual carriers.
\newblock {\em The Journal of Physical Chemistry C}, 119(50):27946--27953,
  2015.

\bibitem{padigi2015potassium}
Prasanna Padigi, Gary Goncher, David Evans, and Raj Solanki.
\newblock Potassium barium hexacyanoferrate--a potential cathode material for
  rechargeable calcium ion batteries.
\newblock {\em Journal of Power Sources}, 273:460--464, 2015.

\bibitem{tojo2016reversible}
Tomohiro Tojo, Yosuke Sugiura, Ryoji Inada, and Yoji Sakurai.
\newblock Reversible calcium ion batteries using a dehydrated prussian blue
  analogue cathode.
\newblock {\em Electrochimica Acta}, 207:22--27, 2016.

\bibitem{lipson2015rechargeable}
Albert~L Lipson, Baofei Pan, Saul~H Lapidus, Chen Liao, John~T Vaughey, and
  Brian~J Ingram.
\newblock Rechargeable ca-ion batteries: a new energy storage system.
\newblock {\em Chemistry of Materials}, 27(24):8442--8447, 2015.

\bibitem{kim2020high}
Sanghyeon Kim, Liang Yin, Myeong~Hwan Lee, Prakash Parajuli, Lauren Blanc,
  Timothy~T. Fister, Haesun Park, Bob~Jin Kwon, Brian~J. Ingram, Peter Zapol,
  Robert~F. Klie, Kisuk Kang, Linda~F. Nazar, Saul~H. Lapidus, and John~T.
  Vaughey.
\newblock High-voltage phosphate cathodes for rechargeable ca-ion batteries.
\newblock {\em ACS Energy Letters}, 5(10):3203--3211, 10 2020.

\bibitem{jeon2020reversible}
Boosik Jeon, Jongwook~W. Heo, Jooeun Hyoung, Hunho~H. Kwak, Dongmin~M. Lee, and
  Seung-Tae Hong.
\newblock Reversible calcium-ion insertion in nasicon-type nav2(po4)3.
\newblock {\em Chemistry of Materials}, 32(20):8772--8780, 10 2020.

\bibitem{xu2021new}
Zheng-Long Xu, Jooha Park, Jian Wang, Hyunseok Moon, Gabin Yoon, Jongwoo Lim,
  Yoon-Joo Ko, Sung-Pyo Cho, Sang-Young Lee, and Kisuk Kang.
\newblock A new high-voltage calcium intercalation host for ultra-stable and
  high-power calcium rechargeable batteries.
\newblock {\em Nature communications}, 12(1):3369, 2021.

\bibitem{tekliye2022exploration}
Dereje~Bekele Tekliye, Ankit Kumar, Xie Weihang, Thelakkattu~Devassy Mercy,
  Pieremanuele Canepa, and Gopalakrishnan Sai~Gautam.
\newblock Exploration of nasicon frameworks as calcium-ion battery electrodes.
\newblock {\em Chemistry of Materials}, 34(22):10133--10143, 2022.

\bibitem{sai2016role}
Gopalakrishnan Sai~Gautam, Pieremanuele Canepa, William~Davidson Richards,
  Rahul Malik, and Gerbrand Ceder.
\newblock Role of structural h2o in intercalation electrodes: the case of mg in
  nanocrystalline xerogel-v2o5.
\newblock {\em Nano Letters}, 16(4):2426--2431, 2016.

\bibitem{hohenberg1964inhomogeneous}
Pierre Hohenberg and Walter Kohn.
\newblock Inhomogeneous electron gas.
\newblock {\em Physical review}, 136(3B):B864, 1964.

\bibitem{kohn1965self}
Walter Kohn and Lu~Jeu Sham.
\newblock Self-consistent equations including exchange and correlation effects.
\newblock {\em Physical review}, 140(4A):A1133, 1965.

\bibitem{blanc2023phase}
Lauren~E. Blanc, Yunyeong Choi, Abhinandan Shyamsunder, Baris Key, Saul~H.
  Lapidus, Chang Li, Liang Yin, Xiang Li, Bharat Gwalani, Yihan Xiao,
  Christopher~J. Bartel, Gerbrand Ceder, and Linda~F. Nazar.
\newblock Phase stability and kinetics of topotactic dual ca2+--na+ ion
  electrochemistry in nasicon nav2(po4)3.
\newblock {\em Chemistry of Materials}, 35(2):468--481, 2023.

\bibitem{bralsford1960effect}
R~Bralsford, PV~Harris, and William~Charles Price.
\newblock The effect of fluorine on the electronic spectra and ionization
  potentials of molecules.
\newblock {\em Proceedings of the Royal Society of London. Series A.
  Mathematical and Physical Sciences}, 258(1295):459--469, 1960.

\bibitem{padhi1998tuning}
AK~Padhi, V~Manivannan, and JB~Goodenough.
\newblock Tuning the position of the redox couples in materials with nasicon
  structure by anionic substitution.
\newblock {\em Journal of the Electrochemical Society}, 145(5):1518, 1998.

\bibitem{hua2021revisiting}
Xiao Hua, Alexander~S Eggeman, Elizabeth Castillo-Mart{\'\i}nez, Rosa Robert,
  Harry~S Geddes, Ziheng Lu, Chris~J Pickard, Wei Meng, Kamila~M Wiaderek,
  Nathalie Pereira, et~al.
\newblock Revisiting metal fluorides as lithium-ion battery cathodes.
\newblock {\em Nature materials}, 20(6):841--850, 2021.

\bibitem{fan2018high}
Xiulin Fan, Enyuan Hu, Xiao Ji, Yizhou Zhu, Fudong Han, Sooyeon Hwang, Jue Liu,
  Seongmin Bak, Zhaohui Ma, Tao Gao, et~al.
\newblock High energy-density and reversibility of iron fluoride cathode
  enabled via an intercalation-extrusion reaction.
\newblock {\em Nature communications}, 9(1):2324, 2018.

\bibitem{ouyang2020effect}
Bin Ouyang, Nongnuch Artrith, Zhengyan Lun, Zinab Jadidi, Daniil~A Kitchaev,
  Huiwen Ji, Alexander Urban, and Gerbrand Ceder.
\newblock Effect of fluorination on lithium transport and short-range order in
  disordered-rocksalt-type lithium-ion battery cathodes.
\newblock {\em Advanced Energy Materials}, 10(10):1903240, 2020.

\bibitem{clement2020cation}
RJ~Cl{\'e}ment, Z~Lun, and G~Ceder.
\newblock Cation-disordered rocksalt transition metal oxides and oxyfluorides
  for high energy lithium-ion cathodes.
\newblock {\em Energy \& Environmental Science}, 13(2):345--373, 2020.

\bibitem{gocheva2009mechanochemical}
Irina~D Gocheva, Manabu Nishijima, Takayuki Doi, Shigeto Okada, Jun-ichi
  Yamaki, and Tetsuaki Nishida.
\newblock Mechanochemical synthesis of namf3 (m= fe, mn, ni) and their
  electrochemical properties as positive electrode materials for sodium
  batteries.
\newblock {\em Journal of Power Sources}, 187(1):247--252, 2009.

\bibitem{kitajou2017cathode}
Ayuko Kitajou, Yuji Ishado, Tomoki Yamashita, Hiroyoshi Momida, Tamio Oguchi,
  and Shigeto Okada.
\newblock Cathode properties of perovskite-type namf3 (m= fe, mn, and co)
  prepared by mechanical ball milling for sodium-ion battery.
\newblock {\em Electrochimica Acta}, 245:424--429, 2017.

\bibitem{dimov2013transition}
Nikolay Dimov, Akihiro Nishimura, Kuniko Chihara, Ayuko Kitajou, Irina~D
  Gocheva, and Shigeto Okada.
\newblock Transition metal namf3 compounds as model systems for studying the
  feasibility of ternary li-mf and na-mf single phases as cathodes for
  lithium--ion and sodium--ion batteries.
\newblock {\em Electrochimica Acta}, 110:214--220, 2013.

\bibitem{sai2020exploring}
Gopalakrishnan Sai~Gautam, Ellen~B Stechel, and Emily~A Carter.
\newblock Exploring ca--ce--m--o (m= 3d transition metal) oxide perovskites for
  solar thermochemical applications.
\newblock {\em Chemistry of Materials}, 32(23):9964--9982, 2020.

\bibitem{wexler2023multiple}
Robert~B Wexler, Gopalakrishnan~Sai Gautam, Robert~T Bell, Sarah Shulda,
  Nicholas~A Strange, Jamie~A Trindell, Joshua~D Sugar, Eli Nygren, Sami
  Sainio, Anthony~H McDaniel, et~al.
\newblock Multiple and nonlocal cation redox in ca--ce--ti--mn oxide
  perovskites for solar thermochemical applications.
\newblock {\em Energy \& Environmental Science}, 16(6):2550--2560, 2023.

\bibitem{euchner2019unlocking}
Holger Euchner, Oliver Clemens, and M~Anji Reddy.
\newblock Unlocking the potential of weberite-type metal fluorides in
  electrochemical energy storage.
\newblock {\em npj Computational Materials}, 5(1):31, 2019.

\bibitem{dey2019topochemical}
Utsav~Kumar Dey, Nabadyuti Barman, Subham Ghosh, Shreya Sarkar, Sebastian~C
  Peter, and Premkumar Senguttuvan.
\newblock Topochemical bottom-up synthesis of 2d-and 3d-sodium iron fluoride
  frameworks.
\newblock {\em Chemistry of Materials}, 31(2):295--299, 2019.

\bibitem{ghosh2024topochemical}
Arindam Ghosh, Dereje~Bekele Tekliye, Emily~E. Foley, Varimalla~Raghavendra
  Reddy, Raphaële~J. Clémenta, Gopalakrishnan Sai~Gautam, and Premkumar
  Senguttuvan.
\newblock Topochemical modulations from 1d iron fluoride precursor to 3d
  frameworks.
\newblock {\em Under review}, 2024.

\bibitem{park2021weberite}
Hyunyoung Park, Yongseok Lee, Min-kyung Cho, Jungmin Kang, Wonseok Ko,
  Young~Hwa Jung, Tae-Yeol Jeon, Jihyun Hong, Hyungsub Kim, Seung-Taek Myung,
  et~al.
\newblock Na 2 fe 2 f 7: a fluoride-based cathode for high power and long life
  na-ion batteries.
\newblock {\em Energy \& Environmental Science}, 14(3):1469--1479, 2021.

\bibitem{shannon1969effective}
RD~T Shannon and C~Tfc Prewitt.
\newblock Effective ionic radii in oxides and fluorides.
\newblock {\em Acta Crystallographica Section B: Structural Crystallography and
  Crystal Chemistry}, 25(5):925--946, 1969.

\bibitem{shannon1976revised}
Robert~D Shannon.
\newblock Revised effective ionic radii and systematic studies of interatomic
  distances in halides and chalcogenides.
\newblock {\em Acta crystallographica section A: crystal physics, diffraction,
  theoretical and general crystallography}, 32(5):751--767, 1976.

\bibitem{knop1982true}
Osvald Knop, T~Stanley Cameron, and Klaus Jochem.
\newblock What is the true space group of weberite?
\newblock {\em Journal of Solid State Chemistry}, 43(2):213--221, 1982.

\bibitem{cai2009complex}
Lu~Cai and Juan~C Nino.
\newblock Complex ceramic structures. i. weberites.
\newblock {\em Acta Crystallographica Section B: Structural Science},
  65(3):269--290, 2009.

\bibitem{yakubovich1993structure}
O~Yakubovich, V~Urusov, W~Massa, G~Frenzen, and D~Babel.
\newblock Structure of na2fe2f7 and structural relations in the family of
  weberites na2miimiiif7.
\newblock {\em Zeitschrift f{\"u}r anorganische und allgemeine Chemie},
  619(11):1909--1919, 1993.

\bibitem{grey2003structural}
IE~Grey, WG~Mumme, TJ~Ness, Robert~S Roth, and KL~Smith.
\newblock Structural relations between weberite and zirconolite
  polytypes—refinements of doped 3t and 4m ca2ta2o7 and 3t cazrti2o7.
\newblock {\em Journal of Solid State Chemistry}, 174(2):285--295, 2003.

\bibitem{shafer1969synthesis}
MW~Shafer.
\newblock The synthesis and characterization of vanadium difluoride, navf3,
  kvf3, and rbvf3.
\newblock {\em Materials Research Bulletin}, 4(12):905--912, 1969.

\bibitem{bernal2020structural}
Fabian~LM Bernal, Jonas Sottmann, David~S Wragg, Helmer Fjellv{\aa}g,
  {\O}ystein~S Fjellv{\aa}g, Christina Drathen, Wojciech~A S{\l}awi{\'n}ski,
  and Ole~Martin L{\o}vvik.
\newblock Structural and magnetic characterization of the elusive jahn-teller
  active nacrf 3.
\newblock {\em Physical Review Materials}, 4(5):054412, 2020.

\bibitem{ratuszna1989structure}
Alicja Ratuszna, Katarzyna Majewska, and Tadeusz Lis.
\newblock Structure of namnf3 at room temperature.
\newblock {\em Acta Crystallographica Section C: Crystal Structure
  Communications}, 45(4):548--551, 1989.

\bibitem{benner1990uber}
G~Benner and R~Hoppe.
\newblock Uber fluoride des zweiwertigen eisens [1]: zum aufbau von nafef3 [1].
\newblock {\em Journal of fluorine chemistry}, 46(2):283--295, 1990.

\bibitem{yusa2012perovskite}
Hitoshi Yusa, Yuichi Shirako, Masaki Akaogi, Hiroshi Kojitani, Naohisa Hirao,
  Yasuo Ohishi, and Takumi Kikegawa.
\newblock Perovskite-to-postperovskite transitions in nanif3 and nacof3 and
  disproportionation of nacof3 postperovskite under high pressure and high
  temperature.
\newblock {\em Inorganic Chemistry}, 51(12):6559--6566, 2012.

\bibitem{sun2016thermodynamic}
Wenhao Sun, Stephen~T Dacek, Shyue~Ping Ong, Geoffroy Hautier, Anubhav Jain,
  William~D Richards, Anthony~C Gamst, Kristin~A Persson, and Gerbrand Ceder.
\newblock The thermodynamic scale of inorganic crystalline metastability.
\newblock {\em Science advances}, 2(11):e1600225, 2016.

\bibitem{tekliye2024accuracy}
Dereje~Bekele Tekliye and Gopalakrishnan~Sai Gautam.
\newblock Accuracy of metagga functionals in describing transition metal
  fluorides.
\newblock {\em arXiv preprint arXiv:2401.10832}, 2024.

\bibitem{hannah2018balance}
Daniel~C Hannah, Gopalakrishnan Sai~Gautam, Pieremanuele Canepa, and Gerbrand
  Ceder.
\newblock On the balance of intercalation and conversion reactions in battery
  cathodes.
\newblock {\em Advanced Energy Materials}, 8(20):1800379, 2018.

\bibitem{sai2015intercalation}
Gopalakrishnan Sai~Gautam, Pieremanuele Canepa, Aziz Abdellahi, Alexander
  Urban, Rahul Malik, and Gerbrand Ceder.
\newblock The intercalation phase diagram of mg in v2o5 from first-principles.
\newblock {\em Chemistry of Materials}, 27(10):3733--3742, 2015.

\bibitem{amatucci1996coo2}
GG~Amatucci, JM~Tarascon, and LC~Klein.
\newblock Coo2, the end member of the li x coo2 solid solution.
\newblock {\em Journal of The Electrochemical Society}, 143(3):1114, 1996.

\bibitem{aurbach2000prototype}
Doron Aurbach, Z~Lu, Alex Schechter, Yossef Gofer, H~Gizbar, R~Turgeman,
  Y~Cohen, Mordechay Moshkovich, and El~Levi.
\newblock Prototype systems for rechargeable magnesium batteries.
\newblock {\em Nature}, 407(6805):724--727, 2000.

\bibitem{malik2013critical}
Rahul Malik, Aziz Abdellahi, and Gerbrand Ceder.
\newblock A critical review of the li insertion mechanisms in lifepo4
  electrodes.
\newblock {\em Journal of the electrochemical society}, 160(5):A3179, 2013.

\bibitem{vargas1996stability}
Rubicelia Vargas and Marcelo Galv{\'a}n.
\newblock On the stability of half-filled shells.
\newblock {\em The Journal of Physical Chemistry}, 100(35):14651--14654, 1996.

\bibitem{shupack1991chemistry}
Saul~I Shupack.
\newblock The chemistry of chromium and some resulting analytical problems.
\newblock {\em Environmental Health Perspectives}, 92:7--11, 1991.

\bibitem{henkelman2000improved}
Graeme Henkelman and Hannes J{\'o}nsson.
\newblock Improved tangent estimate in the nudged elastic band method for
  finding minimum energy paths and saddle points.
\newblock {\em The Journal of chemical physics}, 113(22):9978--9985, 2000.

\bibitem{sheppard2008optimization}
Daniel Sheppard, Rye Terrell, and Graeme Henkelman.
\newblock Optimization methods for finding minimum energy paths.
\newblock {\em The Journal of chemical physics}, 128(13), 2008.

\bibitem{perdew1996generalized}
John~P Perdew, Kieron Burke, and Matthias Ernzerhof.
\newblock Generalized gradient approximation made simple.
\newblock {\em Physical review letters}, 77(18):3865, 1996.

\bibitem{gao2022design}
Yirong Gao, Tara~P Mishra, Shou-Hang Bo, Gopalakrishnan Sai~Gautam, and
  Pieremanuele Canepa.
\newblock Design and characterization of host frameworks for facile magnesium
  transport.
\newblock {\em Annual Review of Materials Research}, 52:129--158, 2022.

\bibitem{devi2022effect}
Reshma Devi, Baltej Singh, Pieremanuele Canepa, and Gopalakrishnan Sai~Gautam.
\newblock Effect of exchange-correlation functionals on the estimation of
  migration barriers in battery materials.
\newblock {\em npj Computational Materials}, 8(1):160, 2022.

\bibitem{long2021assessing}
Olivia~Y Long, Gopalakrishnan~Sai Gautam, and Emily~A Carter.
\newblock Assessing cathode property prediction via exchange-correlation
  functionals with and without long-range dispersion corrections.
\newblock {\em Physical Chemistry Chemical Physics}, 23(43):24726--24737, 2021.

\bibitem{lu2023weberite}
Tenglong Lu, Sheng Meng, and Miao Liu.
\newblock Weberite na $ \_2 $ mm'f $ \_7 $(m, m'= redox-active metal) as
  promising fluoride-based sodium-ion battery cathodes.
\newblock {\em arXiv preprint arXiv:2310.04222}, 2023.

\bibitem{liao2021scalable}
Jiaying Liao, Jingchen Han, Jianzhi Xu, Yichen Du, Yingying Sun, Liping Duan,
  and Xiaosi Zhou.
\newblock Scalable synthesis of na 2 mvf 7 (m= mn, fe, and co) as
  high-performance cathode materials for sodium-ion batteries.
\newblock {\em Chemical Communications}, 57(87):11497--11500, 2021.

\bibitem{kresse1999ultrasoft}
Georg Kresse and Daniel Joubert.
\newblock From ultrasoft pseudopotentials to the projector augmented-wave
  method.
\newblock {\em Physical Review B}, 59(3):1758, 1999.

\bibitem{blochl1994improved}
Peter~E Bl{\"o}chl, Ove Jepsen, and Ole~Krogh Andersen.
\newblock Improved tetrahedron method for brillouin-zone integrations.
\newblock {\em Physical Review B}, 49(23):16223, 1994.

\bibitem{kresse1993ab}
Georg Kresse and J{\"u}rgen Hafner.
\newblock Ab initio molecular dynamics for liquid metals.
\newblock {\em Physical review B}, 47(1):558, 1993.

\bibitem{kresse1996efficient}
Georg Kresse and J{\"u}rgen Furthm{\"u}ller.
\newblock Efficient iterative schemes for ab initio total-energy calculations
  using a plane-wave basis set.
\newblock {\em Physical Review B}, 54(16):11169, 1996.

\bibitem{monkhorst1976special}
Hendrik~J Monkhorst and James~D Pack.
\newblock Special points for brillouin-zone integrations.
\newblock {\em Physical Review B}, 13(12):5188, 1976.

\bibitem{sun2015strongly}
Jianwei Sun, Adrienn Ruzsinszky, and John~P Perdew.
\newblock Strongly constrained and appropriately normed semilocal density
  functional.
\newblock {\em Physical Review Letters}, 115(3):036402, 2015.

\bibitem{dudarev1998electron}
Sergei~L Dudarev, Gianluigi~A Botton, Sergey~Y Savrasov, CJ~Humphreys, and
  Adrian~P Sutton.
\newblock Electron-energy-loss spectra and the structural stability of nickel
  oxide: An {LSDA+\textit{U}} study.
\newblock {\em Physical Review B}, 57(3):1505, 1998.

\bibitem{anisimov1991band}
Vladimir~I Anisimov, Jan Zaanen, and Ole~K Andersen.
\newblock Band theory and mott insulators: Hubbard \textit{U} instead of stoner
  \textit{I}.
\newblock {\em Physical Review B}, 44(3):943, 1991.

\bibitem{gautam2018evaluating}
Gopalakrishnan~Sai Gautam and Emily~A Carter.
\newblock Evaluating transition metal oxides within {DFT-SCAN and
  SCAN+\textit{U}} frameworks for solar thermochemical applications.
\newblock {\em Physical Review Materials}, 2(9):095401, 2018.

\bibitem{long2020evaluating}
Olivia~Y Long, Gopalakrishnan~Sai Gautam, and Emily~A Carter.
\newblock Evaluating optimal \textit{U} for 3\textit{d} transition-metal oxides
  within the {SCAN+\textit{U}} framework.
\newblock {\em Physical Review Materials}, 4(4):045401, 2020.

\bibitem{hellenbrandt2004inorganic}
Mariette Hellenbrandt.
\newblock The inorganic crystal structure database ({ICSD})—present and
  future.
\newblock {\em Crystallography Reviews}, 10(1):17--22, 2004.

\bibitem{laligant1989ordered}
Y~Laligant, Y~Calage, G~Heger, J~Pannetier, and G~Ferey.
\newblock Ordered magnetic frustration: Vii. na2nifef7: Reexamination of its
  crystal structure in the true space group after corrections from renninger
  effect and refinement of its frustrated magnetic structure at 4.2 and 55 k.
\newblock {\em Journal of Solid State Chemistry}, 78(1):66--77, 1989.

\bibitem{chu2018predicting}
Iek-Heng Chu, Sayan Roychowdhury, Daehui Han, Anubhav Jain, and Shyue~Ping Ong.
\newblock Predicting the volumes of crystals.
\newblock {\em Computational Materials Science}, 146:184--192, 2018.

\bibitem{ong2013python}
Shyue~Ping Ong, William~Davidson Richards, Anubhav Jain, Geoffroy Hautier,
  Michael Kocher, Shreyas Cholia, Dan Gunter, Vincent~L Chevrier, Kristin~A
  Persson, and Gerbrand Ceder.
\newblock Python materials genomics (pymatgen): A robust, open-source python
  library for materials analysis.
\newblock {\em Computational Materials Science}, 68:314--319, 2013.

\bibitem{hart2008algorithm}
Gus~LW Hart and Rodney~W Forcade.
\newblock Algorithm for generating derivative structures.
\newblock {\em Physical Review B}, 77(22):224115, 2008.

\bibitem{hart2009generating}
Gus~LW Hart and Rodney~W Forcade.
\newblock Generating derivative structures from multilattices: Algorithm and
  application to {HCP} alloys.
\newblock {\em Physical Review B}, 80(1):014120, 2009.

\bibitem{hart2012generating}
Gus~LW Hart, Lance~J Nelson, and Rodney~W Forcade.
\newblock Generating derivative structures at a fixed concentration.
\newblock {\em Computational Materials Science}, 59:101--107, 2012.

\bibitem{morgan2017generating}
Wiley~S Morgan, Gus~LW Hart, and Rodney~W Forcade.
\newblock Generating derivative superstructures for systems with high
  configurational freedom.
\newblock {\em Computational Materials Science}, 136:144--149, 2017.

\bibitem{zhou2004first}
Fei Zhou, Matteo Cococcioni, Chris~A Marianetti, Dane Morgan, and G~Ceder.
\newblock First-principles prediction of redox potentials in transition-metal
  compounds with lda+ u.
\newblock {\em Physical Review B}, 70(23):235121, 2004.

\end{thebibliography}

\end{document}